# Impact of closing schools on mental health during the COVID-19 pandemic: Evidence using panel data from Japan

Eiji YAMAMURA

Department of Economics, Seinan Gakuin University/ 6-2-92 Nishijin Sawaraku Fukuoka, 814-8511.

Email: yamaei@seinan-gu.ac.jp

Yoshiro TSUSTSUI

Department of Sociology, Kyoto Bunkyo University, Japan.

Email: tsutsui@econ.osaka-u.ac.jp

Corresponding Author: Eiji YAMAMURA. Email: yamaei@seinan-gu.ac.jp

## Abstract

The spread of the novel coronavirus disease caused schools in Japan to close to cope with the pandemic. In response to this, parents of students were obliged to care for their children during the daytime when they were usually at school. Does the increase in burden of childcare influence parents' mental health? Based on short panel data from mid-March to mid-April 2020, we explored how school closures influenced the mental health of parents with school-aged children. Using the fixed effects model, we found that school closures lead to student's mothers suffering from worse mental health than other females, while the fathers' mental health did not differ from other males. This tendency was only observed for less educated mothers who had children attending primary school, but not those attending junior high school. The contribution of this paper is to show that school closures increased the inequality of mental health between genders and the educational background of parents.

## 1. Introduction

To mitigate the coronavirus disease (COVID-19) pandemic, many countries have adopted policies to enforce citizens to stay home in 2020. Under this restricted life, a question that arises: how and to what degree does the COVID-19 outbreak affect mental health? Previous studies have shown that the COVID-19 outbreak has negatively affected mental health (e.g. Brodeur et al. 2020; Sabat et al. 2020; Yamamura and Tsutsui 2020a)[1]. There is a gap in working from home between working mothers having children of primary school age and other working women during the COVID-19 spread (Yamamura and Tsutsui, 2020 b). Changes in working style seem to influence mental health. The allocation of time spent on housework differs between husbands and wives in the normal situation in Japan (Yamamura and Tsutsui 2021). However, it is unknown how school students influence their parents' mental health and how this influence differs between mothers and fathers during the school closure. This study examines the influence of school closure on parents' mental health by focusing on gender differences among parents[2].

The COVID-19 pandemic has drastically changed working styles and time use in various countries[3]. As a consequence of the lockdown to cope with COVID-19, the percentage of people who stay at home increased by 8% across the United States counties (Brzezinski et al. 2020)[4]. In addition, schools were closed because of the emergent situation under the diffusion of COVID-19 in various countries (Baldwin and Mauro 2020). Parents' care for school-aged children plays a critical role in child growth[5]. The closure of primary schools resulted in parents taking care of their children at home, as childcare services were not available because of the COVID-19 pandemic. Therefore, parents' childcare burden increased.

According to the Global Gender Gap Index 2020 rankings, Japan was 121st among 153 countries (World Economic Forum 2020). Under the COVID-19 pandemic, even for two-income households, mainly women worked from home to take care of their primary school children (Yamamura and Tsustui 2020 b)[6]. In Japan, the mental health of mothers with school-aged children was predicted to deteriorate more than fathers' due to school closure caused by the COVID-19 outbreak. The existing research has evaluated the effect of school closure on coping with outbreaks of several viral diseases (e.g. Cauchemez et al. 2008; Cauchemez et al. 2014; Adda 2016) and on parents' working from home (Sevilla and Smith 2020).

---

[1] Existing studies consider the effect of COVID-19 on mental health and subjective view (e.g. Fetzer et al. 2020a, 2020b, Layard et al. 2020).
[2] There were studies that considered the differences in the effects of COVID-19 between genders (Adams 2020; Alon et al. 2020).
[3] Unexpected external shocks, such as the Great Recession, were observed to change time allocation in the daily life (e.g. Aguiar et al. 2013; Gorsuch 2016; Pabilonia 2017).
[4] The recession caused by COVID-19 is different from other types of recessions to the extent that COVID-19 has a greater impact on sectors with high female employment shares (Alon et al. 2020).
[5] Self-care after school increased risk of skipping school and use of alcohol and drugs (Aizer 2004). Economic recessions increased teenagers' risky behaviors (Pabilonia 2017). A mother's absence reduced the time a child spends in school (Pörtner 2016).
[6] One major topic regarding parental time with children in the field of household economics (see Gutiérrez-Domènech 2010; Aguiar et.al. 2013; Gimenez-Nadal and Molina 2014; Morrill and Pabilonia 2015; Gorsuch 2016; Romann and Cortina 2016; Bauer and Sonchak 2017).

Primary and junior high schools were closed throughout Japan from the beginning of March to the end of May 2020 and generally reopened in June 2020. School-aged children could learn at school during the daytime after June 2020[7]. We conducted Internet surveys five times from March to June 2020 to independently construct panel data. Due to the setting and the novel data, we could explore the impact of school closure on parents' mental health by comparing school closure and reopening periods. We found that school closures decreased the mental health of mothers with children who were primary school pupils, which concurs with other COVID-19 work by Takaku and Yokoyama (2020), arguing that marital relationships are deteriorated by school closures. The key finding of this work is that the negative effect of school closures on mental health is observed for lesser educated mothers, but not for higher educated mothers. The contribution of this paper is to show that school closures increased the inequality of parents' mental health between both genders and educational background.

The remainder of this paper is organised as follows. In section 2, we review related literature. Section 3 presents an overview of the situation in Japan. In section 4, we explain the survey design and data. In section 5, the empirical method is described. Section 6 presents and interprets the estimated results. Finally, section 7 provides some reflections and conclusions.

2. Related literature

Under globalization, the severe acute respiratory syndrome coronavirus 2 (SARS-CoV-2) has rapidly diffused all around the world within several months, with many victims[8]. Social networks across countries caused not only the spread of the COVID-19 pandemic, but also spill-overs in perceptions about SARS-CoV-2 risk and social distancing behaviours (Milani 2020). The COVID-19 pandemic drastically changed lifestyles. Strict city lockdowns significantly decreased the SARS-CoV-2 transmission rate in China (Qiu et al. 2020), while the less stringent state of emergency in Japan was also effective in mitigating COVID-19 (Yamamura and Tsutsui 2020a)[9]. In addition to wearing masks and

---

[7] Actually, time schedules of schools were different from previous years. In June and July, generally, schools were operating with staggered attendance to prevent the spread of COVID-19.
[8] Mortality rates from COVID-19 in Lombardia of Italy was knonwn to be remarkably high. However, after controlling for selection bias, mortality rates ranged between 0.1 and 7.5%, which is far smaller than the 17.5% earlier reported (Depalo 2020).
[9] Using a panel dataset covering 80 countries, Ashraf (2020) found that a one standard deviation improvement in

social distancing, the practice of working from home rapidly increased after the spread of SARS-CoV-2 in various countries (e.g. Adams et al. 2020; Bartik et al. 2020; Brynjolfsson et al. 2020; Okubo 2020).

The effect of working from home on productivity depends on the situation. Working from home increased productivity, as observed in experimental studies in China (Bloom et al. 2020). In contrast, working from home reduced the productivity by 60–70% during the COVID-19 pandemic in Japan (Morikawa 2020), although the reduction of productivity of higher educated workers is smaller than lower educated ones[10]. The shift to work from home was considered to have increased average income, however, this also increased income inequality (Bonacini et al. 2020b). Working from home has had impacts not only on productivity and income, but also on other aspects. Working from home increased time living at home, which increased stress.

Childrearing in Japanese households is described by Vogel (1996) where, in the past, mothers were more able to rely on other family members and community members. In contemporary Japan, however, apart from mother and teacher, few other actors are involved in children's lives on a regular basis (Vogel 1996). Husbands have become the most important factor in reducing childrearing stress on mothers (Suzauki et al. 2009), and so researchers have examined the effects of the parental leave policy on fathers' childcare involvement (Ekberg et al. 2013; Kluve and Tamm 2013; Tamm 2019)[11]. Holloway (2010) described the parental role in schooling as follows[12]: Before World War II, fathers rather than mothers played a major role in children's education, particularly in the elite family because mothers were not seen as appropriate tutors for their children. However, the role of mothers changed in the post-war period. There are '…numerous ways in which contemporary Japanese mothers are expected to support their children's academic achievement. They received explicit written instruction from the school on how to reinforce the school routine at home. The schools expect them to help their children maintain the same eating and sleeping schedule, whether they are on vacation or attending school.' (Holloway 2010, pp. 148–149). There is a difference in the burden of mothers' childcare according to her educational background: '…the less educated mothers reported participating in more activities at their child's preschool than do the more highly educated mothers.' (Holloway 2010, p. 153)[13].

School closures were effective in mitigating the spread of COVID-19 in Italy (Bonacini et al. 2020a). In Japan,

---

socioeconomic conditions lowers COVID-19 confirmed cases and deaths per million people by half.

[10] In Japan, the use of telework increased from 6% in January to 10% in March and reached 17% in June 2020 (Okubo 2020).

[11] In Uganda, '…fathers are more likely to engage in childcare when husbands and wives share or have equal wealth than when there are wealth differences between spouses.' (Nkwake 2015, p. 114).

[12] Parents' involvement in a child's education is observed to be important to improve educational attainment (Ordine et al. 2018). The policy of Head Start incentivised parents to spent time in teaching their child (Gelber and Isen 2013).

[13] Holmlund et al. (2011) is an excellent literature review about the relation between parent's schooling and children's schooling.

school closures lead people to stay home even if legally enforced lockdown was announced (Watanabe and Yabu 2020). During the lockdown period in England, mothers spent substantially longer in childcare and housework than fathers (Andrew 2020). 'Compared to their counterparts in the West and in other Asian countries, more Japanese women with children view child rearing as a complex job with few emotional reward.' Only 47% of Japanese parents thought it was '…always enjoyable to raise children, whereas 64 percent of parents in South Korea and 67 percent of parents in the United States responded to this question.' (Holloway 2010, p. 6). Yamaguchi et al. (2018) provided evidence that childcare use improves subjective well-being and reduces stress among low-education mothers in Japan. However, childcare cannot be supplied during emergency situations. The closure of schools critically increases childcare for parents, which disproportionately affects women in Italy during the COVID-19 pandemic (Del Boca et al. 2020). School closures in Japan increased domestic violence, children's weight, and mothers' anxiety over how to raise their children (Takaku and Yokoyama 2020) [14].

In Singapore, the state of lockdown resulted in a clear decline of people's life satisfaction, and its level remained below its level before the COVID-19 spread even after the lockdown was deregulated in Singapore (Cheng et al. 2020). In Japan, Fear and worry about COVID-19 is considered to reflect increased awareness about COVID-19 among people (Sasaki et al. 2020). However, people can feel more happiness even during crisis periods through consumption of cultural goods and services in the past (Tubadij 2020). Fetzer et al. (2020a) provided evidence that higher perceived mortality and contagiousness caused by COVID-19 resulted in increased anxiety about economic outcomes. School closures and social isolation lead school students to become depressed (Asanov et al. 2020). Due to differences in individual characteristics, there is an increase in the gap in mental health among people. For instance, '…mental health in the UK worsened by 8.1% on average as a result of the pandemic and by much more for young adults and for women which are groups that already had lower levels of mental health before COVID-19. Hence, inequalities in mental health have been increased by the pandemic.' (Banks and Xu 2020, p. 91). According to Etheridge and Spantig (2020), in the UK, mental health declined after the COVID-19 pandemic, a decline twice as large for women than for men. One of the factors that explain the gender difference in the decline of mental health is family-related time use and caring responsibilities. However, the impact of COVID-19 changed as time passed and social conditions changed. In Germany, mental health worsened immediately after the lockdown against COVID-19 and started easing with the third week of lockdown (Armbruster and Klotzbuecher 2020).

---

[14] School student's body weight increased after the Great East Japan Earthquake (Yamamura 2016).

## 3. Overview of real situation in Japan

Cases of pneumonia detected in Wuhan, China, were first reported to the World Health Organization (WHO) on 31 December 2019, with the first death observed on 9 January 2020. The first case of a person infected by SARS-CoV-2 was confirmed on 16 January 2020 (the person had previously visited Wuhan). On 28 January 2020, based on the New Infectious Diseases Law, the Government of Japan recognized COVID-19 as a 'designated infectious disease'. Therefore, the Japanese government ordered enforced hospitalisation and restricted work if a person was infected by SARS-CoV-2.

On 27 February 2020, the Japanese government requested schools to close at the beginning of March 2020. The closure of various schools—primary, junior high, and high school—commenced on 2 March 2020. Figure 1 shows the changes in the number of people infected daily with SARS-CoV-2 from 1 March 2020 to 30 June 2020. As shown in Figure 1, the total number of people infected with SARS-CoV-2 was only about 250, and the average number of people infected daily was only 24 at the start of the school closure.

[Insert Figure 1 here]

**Fig. 1** Changes in the number of daily infected persons, the timing of the five waves of the surveys, and the closing and reopening of schools in Japan between 1 March 2020 and 2 July 2020

The number of infected people surged in April 2020. To mitigate the rapid spread of SARS-CoV-2, the government declared a state of emergency on 7 April 2020. As is evident from Figure 1, the third wave of surveys was conducted when the number of daily infected persons peaked. Similar to other countries (Baldwin and Mauro 2020), art museums and amusement parks were closed. However, the government only requested citizens to avoid person-to-person contact and gatherings, encouraging them to stay at home. Even when people did not follow the request, they were neither punished nor penalised under the state of emergency[15]. Therefore, Japanese citizens could behave based on their free will, although moral and informal social norms, to a certain extent, deterred them from practising undesirable behaviour (Yamamura 2009).

During the emergency period, the number of daily infected persons declined remarkably. On 25 May 2020, the state of emergency was deregulated because daily infected persons reduced to approximately 20–30. Naturally, schools were reopened, although reopening varied among regions. All schools reopened throughout Japan by 12 June 2020, when the fifth wave surveys were conducted.

---

[15] The situation was different from countries implementing drastic measures, such as the 'lockdown' in the United States, the United Kingdom, Italy, France, and Spain.

[Insert Figure 2 here]

There are 47 prefectures in Japan. In section 3, we explain how we constructed data on school closure rates in each prefecture and at the time of the survey waves. Based on the data, Figure 2 illustrates a map to show the degree of school closures from Wave 1 to Wave 5. The closure rates of primary schools were almost the same as those of junior high schools. Parts coloured black suggest prefectures where 100% of primary schools were closed, while those coloured white mean prefectures where 0% schools were not closed. Parts heavily and lightly shaded indicate between 90–99% and 1–89%, respectively. At time of Wave 1, with the exception of four prefectures, 100% of primary and junior high schools were closed in most prefectures. However, in Wave 2, the closure rates reduced in rural areas where SARS-CoV-2 hardly spread where total infected persons in these prefectures were smaller than 10. Especially in Iwate, Yamagata, Shimane, and Toyama prefectures, nobody was infected until 27 March 2020, when Wave 2 was conducted. There were 12 prefectures that did not reach 100% school closure. Regarding Wave 3, conducted directly after the declaration of the state of emergency on 7 April 2020, prefectures with 100% school closures increased, which is considered to be influenced by that declaration. In Wave 4, even during the period of the state of emergency, prefectures with closure rates below 100% increased again, although 100% school closures persisted in 30 prefectures. In Wave 5, after deregulation of the state of emergency, the situation drastically changed. There were no prefectures in which the closure rates were over 90%. All schools were completely reopened in 18 prefectures. Further, the degree of reopening schools varied at Wave 5. There were schools that shortened school hours and schools reopened with staggered attendance. If we added these schools to reopening schools, the closure rate decreased to 0% in all prefectures. Therefore, in this case, all parts of Japan can be coloured white. There is a significant gap in the degree of school closures between Wave 5 and other waves.

We began collecting data from 13 March 2020 by conducting the first wave of Internet surveys. During this time, schools were closed throughout Japan. Therefore, households with school-aged children were confronted with unexpected situations of their children not going to school. However, parents could outsource childcare to childcare service providers.

Figures 3 and 4 illustrate the changes over time in mental health, especially in levels of *Anger, Fear, Anxiety*, and *Happiness*, comparing the difference in mental health between individuals with and without primary school children for males (Figure 3) and females (Figure 4). As explained in the next section (and in Table 1 below), higher the *Anger, Fear,* and *Anxiety* levels yielded worse mental health. Meanwhile, higher *Happiness* levels meant happier people.

[Insert Figure 3 here]

**Fig. 3** Mental health comparison of males with and without primary school children on a 5-point scale

Overall, in Figure 3, mental health was stable from the first to the second wave but worsened from the second to the third wave. That is, the mental health of males deteriorated after the declaration of the state of emergency. Subsequently, mental health improved from the third to the fifth waves. A comparison between males with and without primary school children showed no statistical difference in mental health in any of the five waves. Therefore, having primary school children at home did not influence the mental health of males. We interpret this as fathers not taking childcare responsibility even when the children stayed home during the school closure period. This is consistent with the findings of Yamamura and Tsutsui (2020b) that fathers of primary school students were more likely to go to workplaces than working mothers during the school closure. Turning to *Happiness*, males with primary school children are significantly happier than other males. This suggests that fathers enjoyed having children partly because they did not have the stress from childcare. That is, the children had positive (not negative) effects on their fathers. Conversely, happiness levels were almost the same during the study period. That is, changes in mental health were not reflected in happiness levels.

[Insert Figure 4 here]

**Fig. 4** Mental health comparison of females with and without primary school children on a 5-point scale

As shown in Figure 4, similar to males, the declaration of the state of emergency deteriorated the mental health of females. This deterioration peaked immediately after the declaration and improved thereafter. Figures 1, 3, and 4 jointly indicate that mental health was negatively related to the number of daily infected citizens regardless of gender. However, unlike males, the influence of school closure on the mental health of mothers of primary school children was worse than that of other females from the first to the fourth waves, with this difference disappearing during the fifth wave. That is, the mental health of mothers of primary school children improved to the same level as other females once schools reopened. In our interpretation, mothers were burdened with childcare during school closure, which had a negative impact on their mental health[16]. Regarding *Happiness*, there was no significant difference in happiness level between females with primary school children and other females. This means that the negative effect of a small child neutralised its positive effects on mother's happiness. That is, the burden of childcare is large enough to remove the increase in the happiness level from having a child. Meanwhile, the happiness level at Wave 1 was almost the same as that at Wave 5. In comparison, the Happiness level is low in Wave 3, which is directly after the declaration of the state of emergency. That is, changes in mental health were considered to be reflected in happiness levels.

Overall, the observations in Figures 3 and 4 suggest that the stress for mothers with primary school children was larger than that for fathers during the school closure period under the COVID-19 pandemic circumstances.

## 4. Design of surveys and data[17]

### 4.1. Design of surveys

Since the beginning of February 2020, the COVID-19 outbreak spread from China to other countries, although its impact was not significant in Japan. Even before the surge in persons infected by SARS-CoV-2, we planned to conduct surveys to independently collect data to investigate how COVID-19 influenced individual and household behaviours. We commissioned a research company (INTAGE) to conduct surveys through the Internet[18]. A large number of individuals are registered as subjects on the INTAGE list. They vary widely in age, educational background, income level, and others. The sampling method was designed to collect a representative sample of the Japanese population on gender, age, educational background, and location of residence. We collected the data to send questionnaires through the Internet. In the Internet survey, online questionnaires were sent to selected subjects as Japanese citizens aged 16–79 years throughout Japan. The number of subjects depends on budget constraints. We set the starting and ending dates beforehand. Subjects were asked to answer by selecting several choices in which 'unknown' and 'will not answer' are included. Completed surveys were considered as having all questions answered. As illustrated in Figure 1, the surveys were conducted five times from March to June 2020 with the same set of individuals. We pursued the same individuals from the first to the fifth waves with the structure of the data as panel, although some individuals were newly incorporated to maintain the sample size. However, the data used in this study are limited to those included in the sample from the first to the fifth waves. Hence, short-term panel data were constructed.

---

[16] Unfortunately, due to limitation of data, we cannot examine time allocation for childcare.
[17] Yamamura and Tsutsui (2020a, 2020b) used data from the same surveys. However, Yamamura and Tsutsui (2020a) used the sub-sample which was restricted to respondents who resided in nine prefectures where SARS-CoV-2 spread much more rapidly than the other 38 prefectures. Yamamura and Tsutsui (2020b) used the sub-sample, restricted to respondents who were full-time workers. Further, studied period of these works did not cover period of May and June (Waves 4 and 5). Therefore, data used in these works are different from the data used in the present study.
[18] The company is INTAGE, a reputed company with extensive experience in academic research through Internet surveys.

There are a number of advantages of Internet surveys: we can flexibly set the experimental situation and very low cost. Naturally, researchers increasingly used Internet surveys after the 2010s (e.g. Horton et al. 2011; Curces et al. 2013; Kuzimeko et al. 2015). Regarding our study, it was very important to frequently conduct surveys within a short duration because the situation changed rapidly and drastically. It should be noted that the Internet-based survey response pattern may itself have been affected by the pandemic exposure. However, face-to-face surveys were note possibl under the COVID-19 pandemic. A postal survey is also a traditional collection method, however, prospective subjects were unlikely to physically touch and open mail postal questionnaires to avoid infection. In addition, people tend to stay home and were unlikely to reply. The Internet survey enablesd respondents to avoid physical contact and going out. Overall, the Internet survey is more likely to be influenced by COVID-19 than traditional surveys. Hence, we utilised Internet surveys rather than traditional surveys[19].

The first wave was conducted on 13 March 2020. We gathered 4,359 observations and the response rate was 54.7%. In the second wave, the questionnaire was mainly sent to the 4,359 respondents who had completed the first wave. To mitigate the reduction of the sample, a questionnaire was additionally sent to new subjects. Therefore, the response rate reached 80.2% in the second wave because the main subjects tended to have a higher motivation to participate in the surveys. Likewise, were the third, fourth, and fifth wave surveys were conducted. The response rates reached 92.2% (third wave), 91.9% (fourth wave), and 89.4% (fifth wave). The second, third, and fourth waves were conducted in 2020 on 27 March, 10 April, 8 May, and 12 June, respectively.

### 4.2 Data

The total number of observations was 19,740 in the sample covering the first to the fifth waves. As earlier explained in this section, this is a representative sample of the Japanese population. We limited the sample to a sub-sample appropriate to conduct estimations. Parents of primary school pupils were thought to be between 20 and 50 years old, noting female childbearing ages. Hence, we limited the sample to respondents aged between 20–50 years. Further, we limited respondents who completed surveys from the first to the fifth waves to balance the panel data, with the same respondents appearing five times in the sub-sample. In total, 1411 respondents were included (734 males and 677 females) in each wave for the sub-sample, yielding 7055 total observations in the balanced panel data used for estimations.

---

[19] Rapid response research surveys can be done via phone-based interviews. However, Internet surveys are more useful and effective than phone-based interviews because phone-based interviews have many disadvantages: (1) phone numbers are very difficult to obtain because of strict protection of personal data, (2) it is difficult to ask many questions on the phone, (3) many assistants must be employed to interview sufficient number of respondents and put respondents' answers into a dataset file, resulting in larges costs, and (4) assistants could make input errors.

Turning to data of the school closures used for illustrating Figure 2, the government of Japan (Ministry of Education, Culture, Sports, Science and Technology: MEXT) provided interval data on the percentage of school closures for primary and junior high schools. As a whole, in Japan, the primary and junior high school closure rate was were nearly 100% (99%) on 16 March 2020 and continued to be very high at around 95% on 22 April 2020 and 88% on 11 May 2020. However, the rate declined drastically to nearly 0% (1%) on 1 July 2020. MEXT provided closure rates in each prefecture on 22 April 2020 and 11 May 2020[20]. After the state of emergency was deregulated (25 May 2020), there were variations in the degree of reopening, although most schools were reopened[21]. Completely reopening schools amounted to 54%. Schools that shortened school hours accounted for 19%, while schools reopening with staggered attendance reached 26%. We obtained the rate of the complete reopening in each prefecture on 1 July 2020. In addition, a list of scheduled dates of complete reopening was also obtained. Hence, the complete reopening rate in each prefecture on 14 July 2020 (date of Wave 5) can be calculated. Immediately following the school closures, closure rates in each prefecture were not provided. However, a list of schools that were not closed on 16 March 2020 was available[22]. From the list, we obtained a schedule for the reopening date of schools in several cities and towns. We aggregated the data of cities and towns to obtain the number of school openings in each prefecture. Then, we used this data and the total number of schools from the official data to calculate the school closure rate in each prefecture even when the rate was not available[23]. Thus we construct the prefecture-level panel data of school closure rates for 47 prefectures in the five waves.

Schools that shortened school hours and those that staggered attendance was considered a reopening school. Based on the data where the school closure rate was 0 in Wave 5, the primary school closure rate (*School Closure A*) is defined. However, there was another way to calculate the school closure rate. Here, the estimated closure rate in Wave 5 was used, although the school closure rate was the same as *School Closure A* in Waves 1–4. Thereafter, we constructed the alternative school closure rate, which is defined as *School Closure B*. For this reason, Table 1 indicates that *School Closure A* is lower than *School Closure B*.

---

[20] Source of data for 22 April 2020: https://resemom.jp/article/2020/04/27/56010.html; Source of data for 11 May 2020: https://resemom.jp/article/2020/05/14/56234.html. Accessed on 30 October 2020.
[21] Source of data for 1 July 2020: https://www.mext.go.jp/content/20200603-mxt_kouhou01-000004520_4.pdf. Accessed on 30 October 2020.
[22] Source of data for 16 March 2020: https://www.mext.go.jp/content/20200323-mxt_kouhou01-000006011_13.pdf. Accessed on 30 October 2020.)
[23] We obtained the total number of primary schools from the MEXT Statistics Handbook, 2012.

[Insert Table 1 here]

The descriptions of the variables used in this study are presented in Table 1. The mean values of *Anger, Fear,* and *Anxiety* for females were larger than those for males, indicating that mental health was worse for females than for males. The survey questionnaire contained basic questions about demographics, such as age, gender, educational background, and household income, and about having children in primary or junior high school. We assumed that these variables did not change because the five waves were conducted within a short period. To examine the effect of school-aged children, we made dummy variables for primary school pupils, *primary students*, and junior high school students, *junior high*. It is plausible that childcare is more important for less mature children. Therefore, *primary school*, a dummy variable for having a child in primary school, is the key independent variable.

The Big Five Model is the most widely accepted personality theory not only in the field of social psychology, but also in other fields such as personal economics (Uysal and Pohlmeier 2011). According to the theory, personality can be boiled down to five core factors such as neuroticism, conscientiousness, agreeableness, openness to experience, and extraversion (Norman 1963). The five factors are related to economic outcomes, such as duration of unemployment (Uysal and Pohlmeier 2011). Among these five factors, neuroticism (including 'Anger', 'Anxiety', and 'Fear') changed through experience in the life (Jeronimus et al, 2014). Under budget constraints, we put focus on 'Anger', 'Anxiety', 'Fear' to examine the impact of school closure on neuroticism of parents of students. Apart from neuroticism, a number of studies have examined the impact of disastrous events on subjective wellbeing (Carroll et al. 2009; Luechinger and Saschkly 2009). COVID-19 is a kind of unexpected disastrous event. Therefore, it is valuable to explore the relationship between COVID-19 and subjective wellbeing. Thereafter, we consider the relationship between neuroticism and subjective well-being, which is also a major topic in social science (Cohn et al. 2009; Jibeen 2014).

In Waves 1–5, respondents were asked the following questions:

*'Within two weeks, how much have you felt the emotions of anger, fear, and anxiety? Please answer on a scale of 1 (I have not felt this emotion at all) to 5 (I have felt this emotion strongly).'*

(1) Anger

(2) Fear

(3) Anxiety

The answers to these questions were proxies for mental health: *Anger, Fear, and Anxiety*. In this study, mental health is considered to be worse the larger these values.

These values partially reflect a negative feeling, which is thought to depend on how respondents were questioned. It is necessary to consider positive and global measures of subjective well-being. Therefore, we used the happiness level. Hence, the surveys included the following questions:

*'Now, to what degree are you currently feeling happiness? On a scale of 0 to 10, where 10 is "very happy" and 0 is "very unhappy," how do you rate your current level?'*.

The answer to it was a proxy for subjective wellbeing: *Happiness*. Unlike proxies for mental health, people feel happier when *Happiness* is greater.

## 5. Hypothesis and Methodology

### 5.1 Hypothesis

It is necessary to care for less mature children because they are less able to be independent and self-supporting. Usually, in Japan, there is public support to childcare via childcare centres. In addition, childcare is supplied in the market. However, during the period of COVID-19, especially under the state of emergency, the childcare centre services were drastically reduced to avoid infection. Inevitably, school closures increase parents' time spent on childcare within a household if the child is a primary school student. Further, because of women's gender identity (Akerlof and Kranton 2000), an increase in the burden of childcare might be observed for mothers yet not for fathers. Therefore, the mother is subject to great stress. Accordingly, we proposed hypothesis 1.

*Hypothesis 1: Closure of primary school worsened mother's mental health.*

Matured children are able to do housework and help their mothers. Possibly, being adolescents, the presence of secondary school kids can complement mothers' time input at home in ways that pre-adolescent primary school children cannot. Therefore, closures of junior high schools possibly increased children's time spent on housework, thus decreased mothers'. Accordingly, we proposed hypothesis 2.

*Hypothesis 2: Closures of junior high schools improve a mother's mental health.*

It would be interesting to investigate the relationship between how long (prior to 2 June 2020) the school had reopened and mental health. However, unfortunately, we could not obtain this information. Hence, we cannot examine this relation.

**5.2 Fixed effects model**

The estimated function takes the following form:

$$Y_{it} = a_1 Wave1_t \times Primary_i + a_2 Wave2_t \times Primary_i + a_3 Wave4_t \times Primary_i + a_4 Wave5_t \times Primary_i + a_5 Wave1_t \times Junior\ High_i + a_6 Wave2_t \times Junior\ High_i + a_7 Wave4_t \times Junior\ High_i + a_8 Wave5_t \times Junior\ High_i + \alpha_9 Wave1_t + \alpha_{10} Wave2_t + \alpha_{11} Wave4_t + \alpha_{12} Wave5_t + k_i + u_{it},$$

where $Y_{it}$ represents the dependent variable for individuals $i$ in Wave $t$. The regression parameters and the error term are denoted by $\alpha$ and $u$, respectively. The dependent variables, *Anger, Fear, Anxiety*, and *Happiness* differ according to specifications. $k_i$ is the time-invariant characteristics of the respondents. The fixed effects method was used to control for various time-invariant variables. The fixed effects controlled for all variables have the same value during the studied period between March and July, such as *Schooling, Income, and Age*[24]. Furthermore, *Primary*$_i$ and *Junior High*$_i$ were also controlled, which means that the results cannot be calculated for *Primary*$_i$ and *Junior High*[25]. The critical issue is to explore the impact of school closures on the mental health and happiness of parents of primary school students.

The situation in Japan drastically changed during the study period, as illustrated in Figure 1. In addition, Figure 2 revealed that the degree of school closure was different between the Waves. Even using the fixed effects model, we can calculate how the influence of *Primary* changed between waves by using the cross term between *Primary* and wave dummies: *Primary*$_i$×*Wave1*$_t$, *Primary*$_i$×*Wave2*, *Primary*$_i$×*Wave4*, and *Primary*$_i$×*Wave5*. This also holds between *Junior High* and wave dummies. Hence, these cross-terms were included in the estimation. Time-point-specific effects were controlled by wave dummies. As shown in Figure 1, the state of emergency was declared just before Wave 3, and so

---
[24] Annual household income was asked in the first wave, but not in the other waves. So, in this paper, income level is assumed to be the same during the studied period.
[25] It is possible that primary school students entered junior high school in April if they were in sixth grade in March under the educational system of Japan. Similarly, junior high school students possibly entered junior high school in April if they were in third grade in March. However, we only asked respondents whether they have a child in primary school (junior high school) in *Wave 1*. Thus, we assume that *Primary* and *Junior High* were constant from *Wave 1* to *Wave 5*.

Japanese people were most likely to stay home (Yamamura and Tsutsui 2020a), and the stress on parents is thought to be very high. Hence, we set Wave 3 as the reference category. Therefore, the second (*Wave 1*), third (*Wave 2*), fourth (*Wave 4*), and fifth (*Wave 5*) wave dummies are included, with their reference group being the third wave. Further, cross-terms between *Primary* and wave dummies can be interpreted as suggesting how mental health is different from that at the third wave when COVID-19 dominantly spread. The key variables are cross-terms between *Primary* and wave dummies, especially *the Primary$_i$ ×Wave5*, to test *Hypothesis 1*. Based on *Hypothesis 1*, the signs of these coefficients were predicted to be negative, which means that the mental health of parents with students is improved at Wave 5, compared to Wave 3. From *Hypothesis 2*, the cross-terms between *Junior High* and wave dummies were predicted to have a positive sign. Further, we aim to determine the difference in the childcare burden between parents. Examining Figures 3 to 4, we can expect the reopening of primary schools to improve the mental health of women with primary school children. For a closer examination, by using the fixed effects model, we divided the sample into males and females to conduct the model for gender comparison.

Macro-economic shocks have a negative impact on society throughout Japan at the same time, even though the school closure rate varied according to prefectures at each time point, as shown in Figure 2. We cannot disentangle school closure effects from other time-specific effects when we use wave dummies. For closer examination, it is necessary to identify the school to which a child belongs and then obtain information on whether the school was closed at each time point when we conducted the survey. Unfortunately, we do not have such precise data. We only know the prefecture and its school closure rate in each time period. Using this data, however, we know the probability that the school to which respondent's child belonged was closed in each time period. Hence, in the alternative model, we used the school closure rate as a key variable. The estimated function of the alternative model is

$Y_{it} = b_1 School\ Closure_{it} \times Primary_i + b_2\ School\ Cosure_{it} \times Junior\ High_i + b_3\ School\ Closure_{it} + b_4 Wave1_t + b_5 Wave\ 2_t + b_6 Wave\ 4_t + b_7 Wave\ 5_t + k_i + e_{it}.$

To test this hypothesis, we include cross-terms such as *School Slosure$_{it}$ ×Primary$_i$* and *School closure$_{it}$ ×Junior High*. We predicted signs of these coefficients to be positive, meaning that school closure has a detrimental effect on mental health. Thus, as *School Closure,* we can use *School Closure A*. In addition, as an alternative model, *School Closure B* is used for a robustness check.

Dependent variables are discrete and ordered values. Hence, for a robustness check, we also conducted estimations using the random effects of the ordered probit model for baseline estimations.

# 6 Results and interpretation

## 6.1 Results of cross-terms with wave dummies

[Insert Table 2 here]

Tables 2, 3, and 4 report the results of the fixed-effects model with cross-terms between *Primary* (*Junior High*) and wave dummies. The sample on which the results of Table 2 are based is those respondents equal to or below 50 years old, because we limited respondents who were possibly parents of primary school students. Further, we also used the sub-sample to closely examine how results differ according to educational background. We divided the sample into a sub-sample of respondents whose schooling years were equal to or over 16 (i.e. graduated from university at least) and those below 16 years, respectively. The former is defined as a sub-sample of a high-educated sub-sample of high-educated persons, while the latter is defined as that of low-educated persons. Table 3 shows the results using the sub-sample of high-educated persons, and Table 4 shows the results of the sub-sample of low-educated ones. In these tables, the right and left parts indicate the results of males and females, respectively.

Table 2 suggests that cross-terms are not statistically significant using the male sample. However, using the female sample, *Wave5 ×Primary* shows a negative sign and statistical significance in columns (5)–(7), meaning that female mental health improved after reopening schools than the most serious period, Wave 3. The absolute values of the coefficients of *Wave5 ×Primary* are 0.18, 0.20, and 0.23 for *Anger, Fear,* and *Anxiety*, respectively. Compared to the third wave, mental health values improved by approximately 0.2 on the five-point scale in the fifth wave. These results are consistent with *Hypothesis 1*. However, the happiness level is not influenced by any cross-terms. Interestingly, cross-terms between *Junior High* and wave dummies are positive and statistically significant in column (6), meaning that mothers of junior high school students were less likely to suffer Fear in the strict school closure time (Wave 3) than other periods. This is in line with *Hypothesis 2*. However, these cross-terms are not statistically significant in most cases in other estimations of mental health. Therefore, *Hypothesis 2* is partially supported. Concerning wave dummies, most of the results show statistical significance in all columns. Further, the coefficients of *Anger, Fear,* and *Anxiety* indicate negative signs, while *Happiness* shows positive signs. Considering the results of Tables 2 and 11 and Figure 1 together, we argue that the time point at which the spread of COVID19 peaked resulted in the worst mental health states.

For a robustness check, the results of the random effects of the ordered probit model are shown in Appendix, Table

11. The findings in Table 11 did not change from Table 2. Hence, the results in Table 2 do not depend on the estimation method and are robust to alternative estimation.

[Insert Table 3 here]

In Table 3, most of the cross-terms did not show statistical significance. However, it is interesting to observe that *Wave5 ×Primary* indicates a negative sign and statistically significant in estimations of *Fear* and *Anxiety* for males, but not for females. Its absolute values of coefficients are 0.20, and 0.23 for *Fear* and *Anxiety*, respectively. These values are almost the same as those of females in Table 2. In our interpretation, highly educated fathers were more likely to bear the burden of childcare, and so their mental health was better after the reopening of schools than directly after the declaration of the state of emergency. One interpretation is that high-educated males are more likely to have progressive views and behaviors (Oswald and Powdthavee 2010). Hence, high-educated males participate in housework and childcare during the COVIDO-19 (Del Boca et al. 2020). Assuming that the wife of a highly educated husband is also highly educated, the mother's burden of childcare is smaller because her husband spends his time on housework. This is reflected in the fact that the amount of mother's childcare is not so influenced by school closure. This is in line with the argument of Suzuki et al. (2009) that husbands played an important role in reducing the stress related to childrearing on mothers.

[Insert Table 4 here]

Turning to the results of the low-educated respondents, Table 4 shows that the results of the cross-terms are similar to those in Table 2 when the dependent variable is *Anger, Fear,* and *Anxiety*. This implies that low-educated mothers spent a larger amount of time on childcare during the school closure period. The absolute values of the coefficients of *Wave5 ×Primary* are 0.21, 0.30, and 0.30 for *Anger, Fear,* and *Anxiety*, respectively. These values are larger than those of Table 2 arguably because the gap in the amount of childcare between wife and husband is larger for lower educated couples. As for the results of the estimation of *Happiness*, *Wave1 × primary* and *Wave2 × primary* shows a positive sign and statistical significance, whereas *Wave4 ×Primary* and *Wave5 ×Primary* are not statistically significant. This shows that the mother of the child of primary school students was happier before the declaration of the state of emergency, compared with directly after the declaration. However, her happiness level after reopening school was not higher than that directly after the declaration. In our interpretation, happiness level is unlikely to be influenced by mental health.

**6.2 Results of cross-terms with closure rates.**

Tables 5 and 6 report the results of the fixed-effects model with cross-terms between *Primary* (*Junior High*) and school closure rates using male and female samples, respectively. Further, results using low- and high-educated respondents are shown in Tables 7–8 and 9–10, respectively. In columns (1)–(4), the results of Specification A where *School Closure A* is used. Meanwhile, in columns (5)–(8), the results of Specification B where *School Closure B* is utilised.

[Insert Tables 5 and 6 here]

Table 5 shows no statistical significance in any columns when we checked the cross-terms. This implies that the child did not influence the male's mental health during the school closure period. In contrast, Table 6 indicates the significant positive signs of *School Closure A ×Primary* and *School Closure B ×Primary*, with the exception of *Happiness* as the dependent variable. This implies that school closure has a detrimental effect on the mental health of mothers having children of primary school students. Meanwhile, the statistical insignificance of *Happiness* is consistent with the argument of Cohn et al. (2009) that negative emotions had weak or null effects and did not interfere with positive emotions. However, as a whole, the results are in line with *Hypothesis 1*. There is no statistical significance for *School Closure A ×Junior High* and *School Closure B ×Junior High*. Hence, contrary to Table 2, *Hypothesis 2* was not supported. For a robustness check, the results of the random effects of the ordered probit model are shown in the Appendix, Tables 12 and 13. The findings in Tables 12 and 13 are similar to those of Tables 5 and 6, showing that the results of Tables 5 and 6 are robust to alternative estimation methods.

[Insert Tables 7 and 8 here]

As for the results of highly educated males, Table 7 indicates a significant positive sign of *School Closure A ×Primary* and *School Closure B ×Primary* in the estimation of *Anxiety,* although this tendency is not observed for *Anger* and *Fear*. Considering Tables 3 and 7 jointly means that school closure partly deteriorates the mental health of highly educated males having children of primary school students. In Table 8, a significant positive sign is observed for *School Closure B ×Junior High,* but not for *School Closure A ×Junior High* when *Anger* is the dependent variable. Other results do not show statistical significance. As a whole, school closure did not influence mental health and happiness for highly educated mothers.

[Insert Tables 9 and 10 here]

Turning to the low-educated sample, in Table 9, any statistical significance is observed for cross-terms. Hence, low-educated males having children of school students are not influenced by school closure. Switching our attention to Table 10, both *School Closure A ×Primary* and *School Closure B ×Primary* shows a significant positive sign for the estimation of *Anger, Fear,* and *Anxiety.* The combined results of Tables 4 and 10 reveal that school closure deteriorates the mental health of females having children of primary school students. *Hypothesis 1* is strongly supported. On the other hand, *School Closure A ×Junior High* and *School Closure B ×Junior High* are not statistically significant, which does not support Hypothesis 2.

**6.3 Discussion**

COVID-19 has various detrimental effects on society from various angles, and we cannot know the time when COVID-19 will be eradicated. It is important to consider health policies to cope with the COVID-19 pandemic to sustain society (Bal et al. 2020). The gender gap is larger in Japan than in other developed countries (Brinton 1993; Nemoto 2016). According to Yamamura and Tsutsui (2020b), as a consequence of school closures in Japan, female workers with children at primary school are more likely to work from home. However, this does not hold for female workers having students of junior high school and male workers with children, regardless of the type of school. Accordingly, working women with small children also work as caregivers for their children (Akerlof and Kranton 2000). In the case where the wife is a houseworker while her husband is a worker, because of division of labour within a household, the wife is likely to care for their child during school closures. Yamamura and Tsutsui (2020a) found that people's mental condition worsened with the spread of COVID-19 in Japan. Particularly, within a household, the gap in the childcare responsibility between the mother and father may have a negative influence on the mental health of the former. Under the novel setting of unexpected school closures due to the COVID-19 pandemic, the gender gap in childcare was demonstrated to increase.

Our findings are based on the assumption that mothers are more obliged to care for children. However, this assumption is inconsistent with findings in the United Kingdom that the gender childcare gap has reduced amidst the COVID-19 pandemic (Sevilla and Smith 2020)[26]. The differences in findings between the United Kingdom and Japan can be interpreted in various ways. Women's social status in Japan is far lower than that of women in other developed Western countries such as the United Kingdom (Nemoto 2016). Therefore because of differences in bargaining power, mothers are more obliged to shoulder a larger share of childcare responsibilities (Yamamura and Tsutsui 2021).

---
[26] Sevilla and Smith (2020) analysed gender allocation of childcare within couples with children aged less than 12 years in the United Kingdom.

Due to a dearth of childcare in the market under the COVID-19 in Japan, especially during the state of emergency, mothers are more likely to spend time for childcare by themselves, regardless of their educational background. We found that mothers with low education levels suffer from a larger negative effect of COVID19 than highly educated mothers. One possible interpretation is as follows: the low-educated mothers are less able to devise a way to enjoy spending time with small children, partly because they are less likely to gather appropriate and useful knowledge about childcare during the emergent situation when they cannot go out with their child. If so, a policy is required to support a low-educated mother to learn how to spend time with their child during their stay at home.

Under the school closures, marital relationships worsened in March 2020, but did not persist for four months until August 2020 (Takaku and Yokoyama 2020). Thus, this kind of negative effect of school closure is temporary if the school is reopened. However, the experience of school closure possibly dented parenting confidence. COVID-19 disturbed the rhythm of the child's life, which increased the children's body weight and hence the mother felt anxious about childrearing, which persisted until August (Takaku and Yokoyama 2020). Researchers view the 'mothers' lack of parenting confidence…as one of the most serious problems facing families in contemporary.' (Holloway 2010, p. 7). Policymakers should consider a care mechanism for the mental health of mothers, especially those with small children. Furthermore, by bridging health and labour issues, we should enhance fathers' childcare responsibilities toward small children by promoting working remotely from home under the COVID-19 pandemic. From a long-term and broad perspective, it is critical to consider how to maintain work–life balance and reduce the childcare responsibility gap between husband and wife in designing the post-COVID-19 society.

# 7    Conclusion

School closures mitigate the spread of COVID-19 (Bonacini et al. 2020a). However, the burden on mothers for childcare increased more than that of fathers (Andrew 2020). This study examined whether the mental health of mothers of school students was more likely to worsen than of fathers. The major findings are that school closures lead mothers with primary school children to be in worse mental condition than other females, although this difference disappeared after the reopening of schools. Further, negative effects of school closure were observed only for low-educated parents with educational backgrounds who graduated from high school or less. Meanwhile, we hardly observed the detrimental effects of school closures on fathers with primary school children and parents of junior high schools.

The contribution of this work is that school closures have increased the gap in parents' mental health between their genders and also between educational background. However, school closures do not influence the happiness of parents. A possible interpretation is that the effect of school closure on mother's happiness is neutralized if gender identity increases women's utility from childcare (Akerlof and Kranton 2000).

It is valuable to scrutinise how and why the influence of school closures on mental health differs from that on happiness. Further, the assumption of this study is contrary to findings in the United Kingdom that the gender childcare gap has reduced amidst the COVID-19 pandemic (Sevilla and Smith 2020)[27]. It is critical to investigate how the time allocation for childcare changed during the COVID-19 spread. However because of insufficient data, we could not examine changes in parents' time allocation after school closure. These remain issues to be addressed in future research.

**Declarations**

Funding: Fostering Joint International Research B (Grant No.18KK0048): TSUTSUI received from the Japan Society for the Promotion of Science.

Conflicts of interest: There are no conflicts of interest to declare.

Availability of data and material: Available upon request from the corresponding author.

Code availability: Available upon request for the corresponding author.

**Acknowledgement:** We would like to thank the four anonymous referees and the editor, Klaus F. Zimmermann for their useful suggestions. We also thank Editage (www.editage.jp) for English language editing. This study was supported by Fostering Joint International Research B (Grant No.18KK0048) from the Japan Society for the Promotion of Science.

---

[27] Sevilla and Smith (2020) analysed gender allocation of childcare within couples with children aged less than 12 years in the United Kingdom.

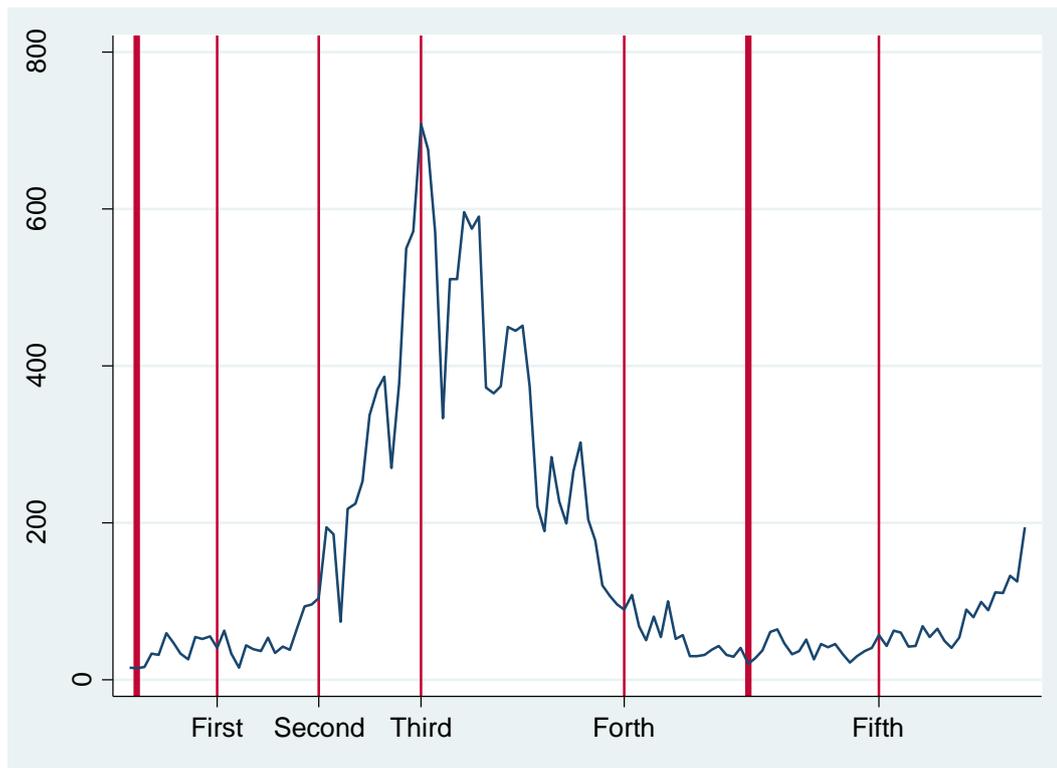

Figure 1. Changes of total infected persons in Japan between 1 March and 18 April 2020, with the timing of the surveys and closing and reopening schools.

Note: First, second, third, fourth, and fifth waves were conducted in 2020 on 10 March, 27 March, 10 April, 8 May, and 12 June, respectively. Thin lines show the dates of the surveys. Thick lines show the date when school closures begun (2 March 2020) and the date when the state of emergency in Japan was deregulated (25 May 2020). After the deregulation, schools were reopened, although actual reopening varied according to prefecture. The state of emergency was promulgated from 7 April 2020 its date is indicated in the Figure.

Source: Details of persons daily infected by COVID 19 were sourced from the official site of the Ministry of Health, Labour and Welfare. https://www.mhlw.go.jp/stf/covid-19/open-data.html. (Accessed 4 July 2020).

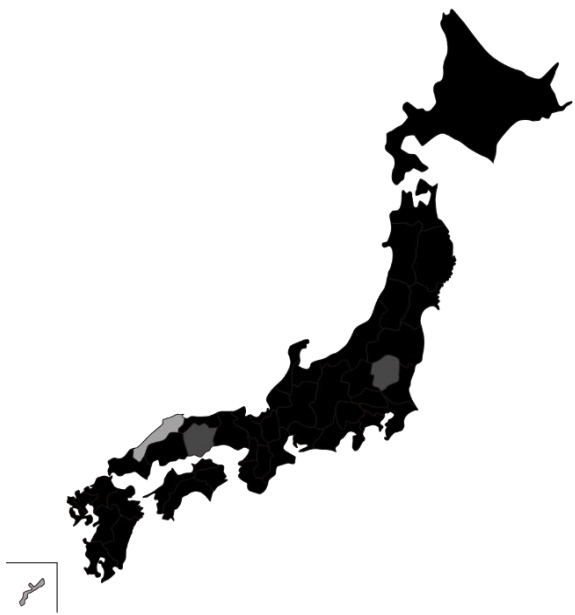

Wave 1

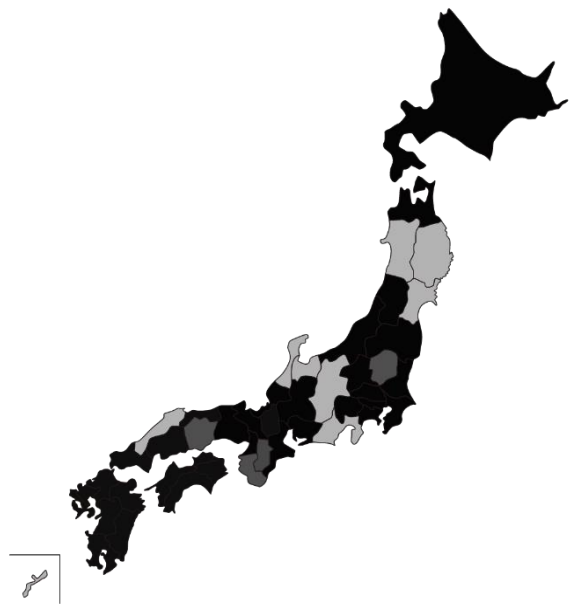

Wave 2

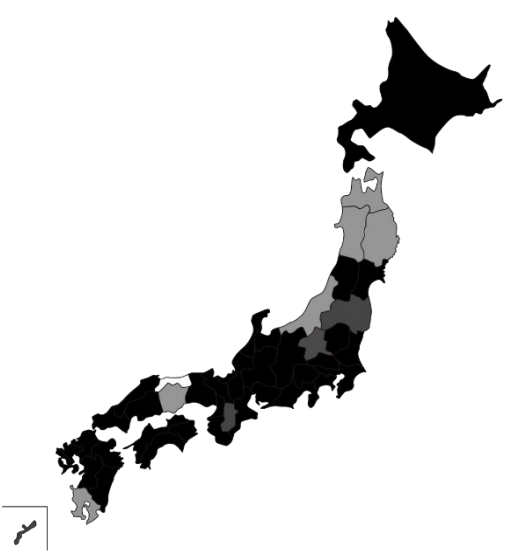

Wave 3

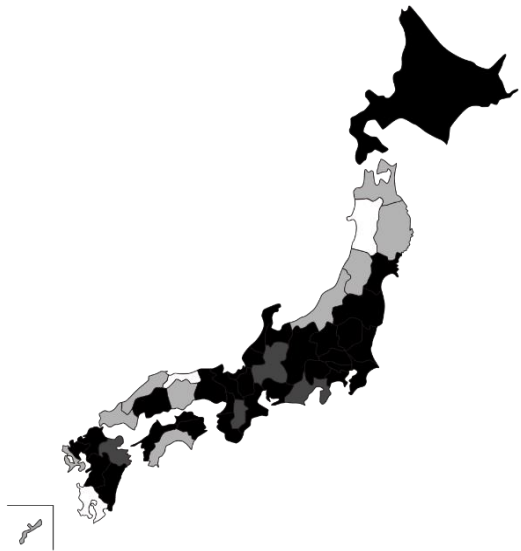

Wave 4

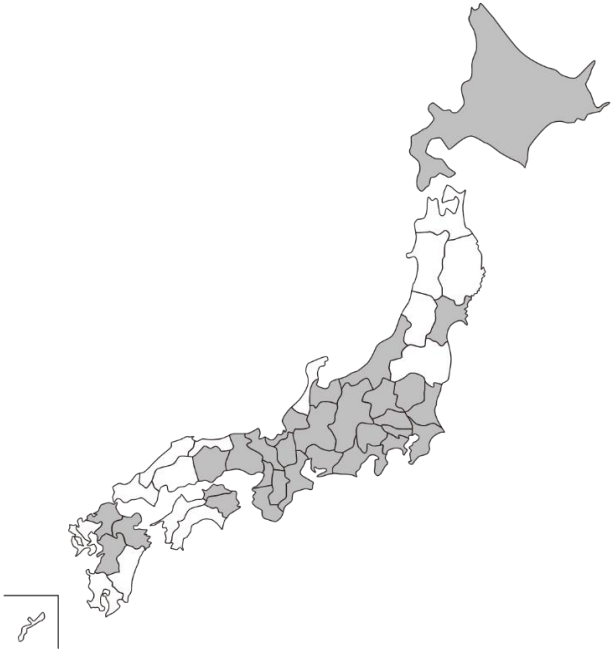

Wave 5

Figure 2. Variations of school closure rates according to prefectures in each period.

Notes: Prefectures coloured black indicate where 100% of primary schools were closed, while those coloured white indicate where 0% schools remained open. Prefectures shaded dark and light indicate between 90–99% and 1–89%, respectively. In Wave 5, all prefectures were coloured white if the partial reopening was not considered as closed.

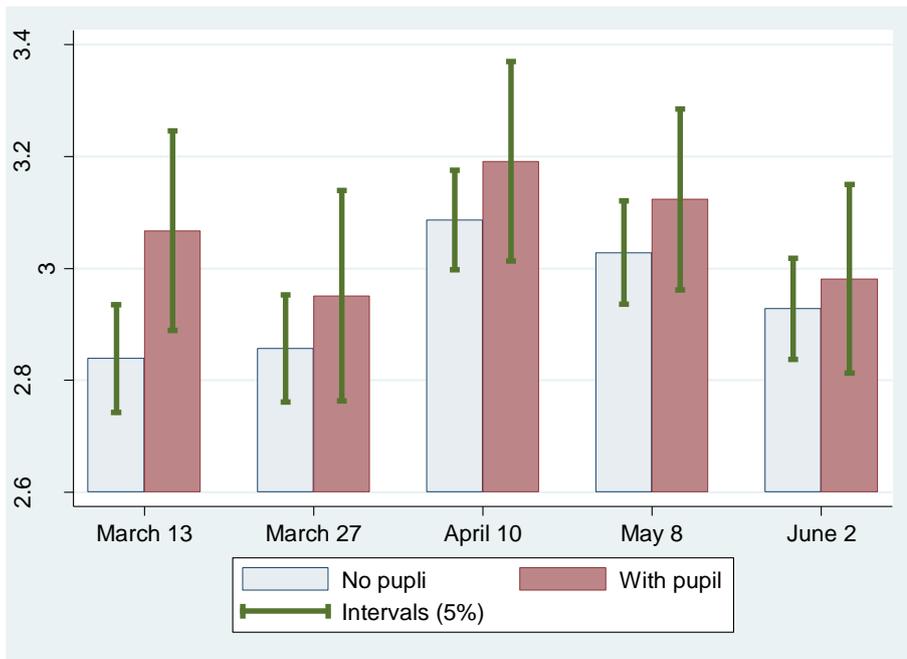

Anger

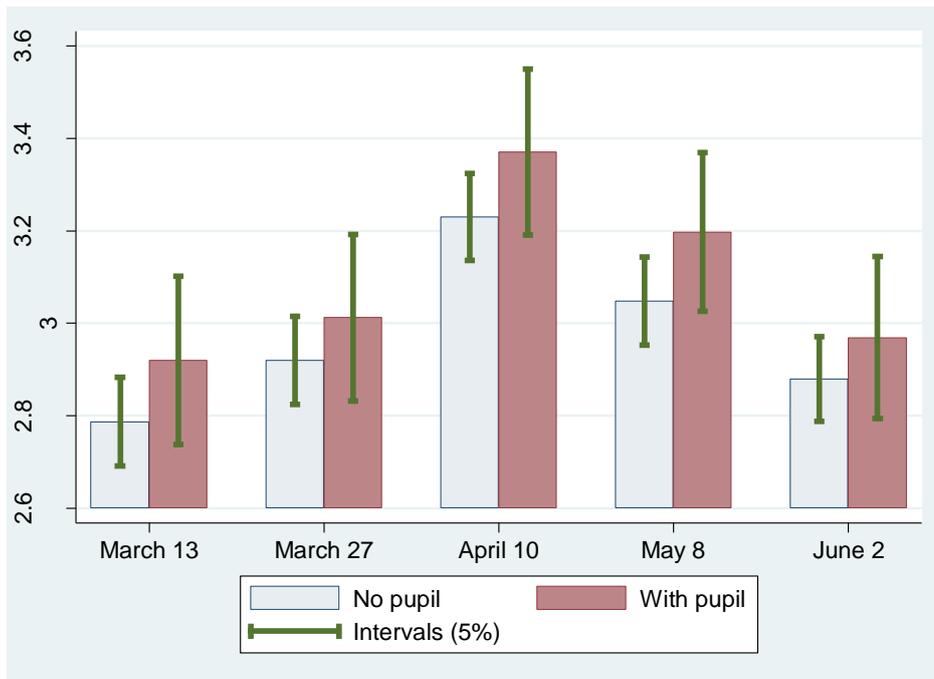

Fear

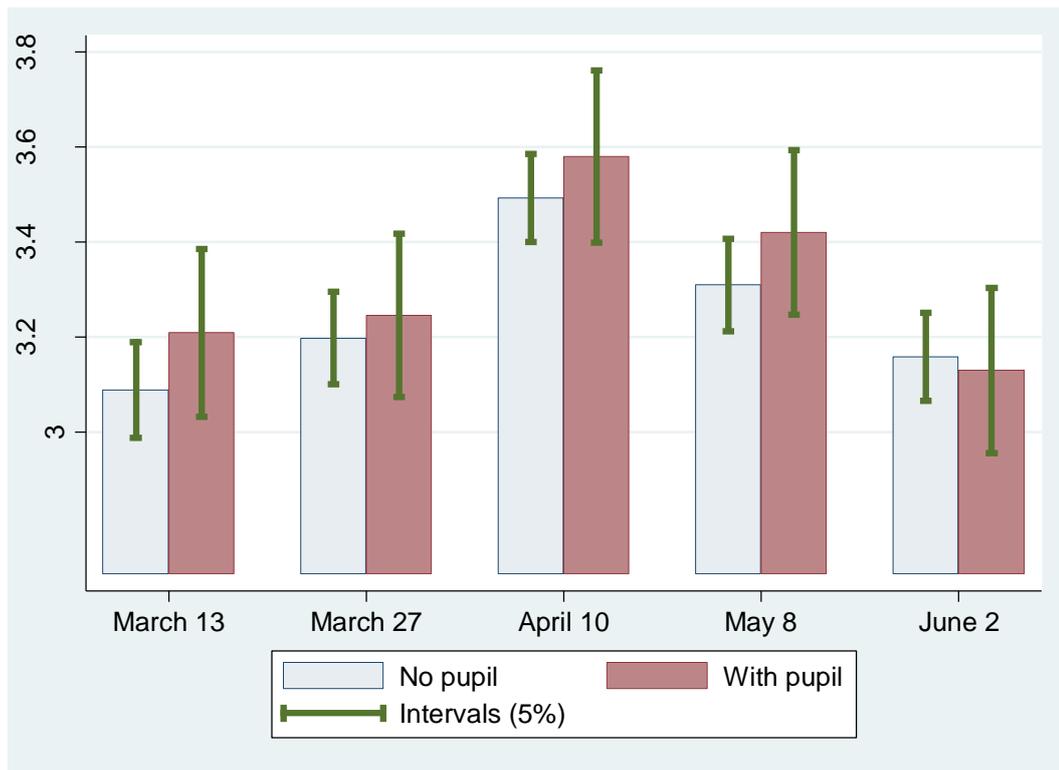

Anxiety

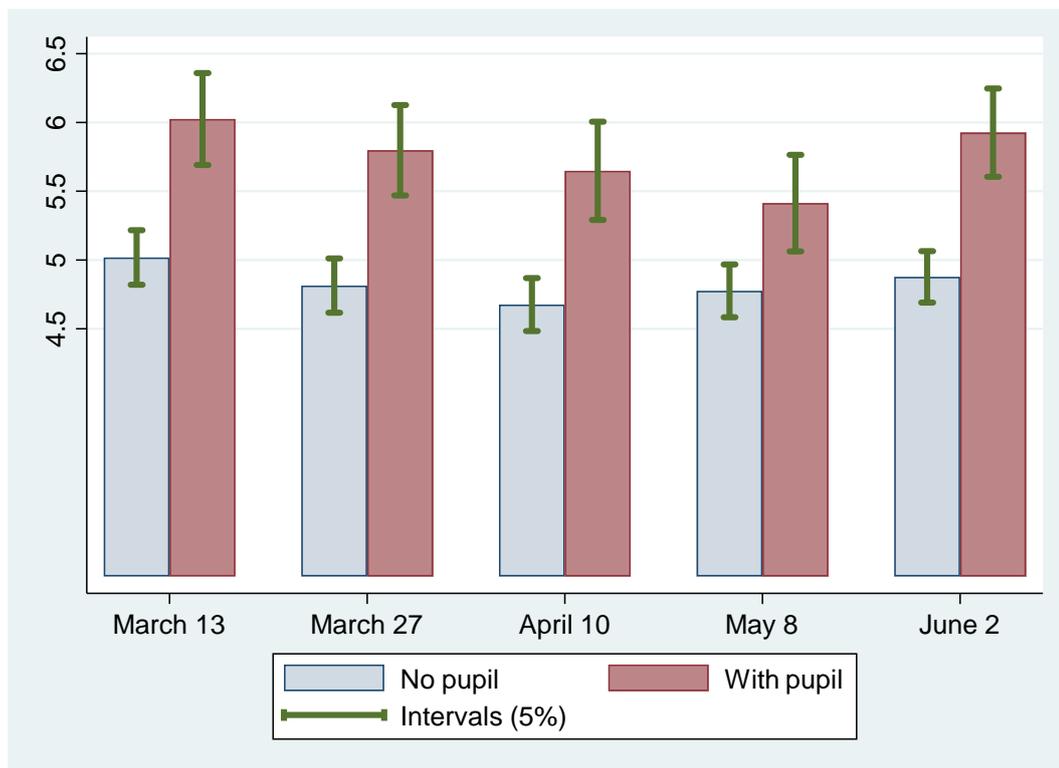

Happiness

Figure 3. Male

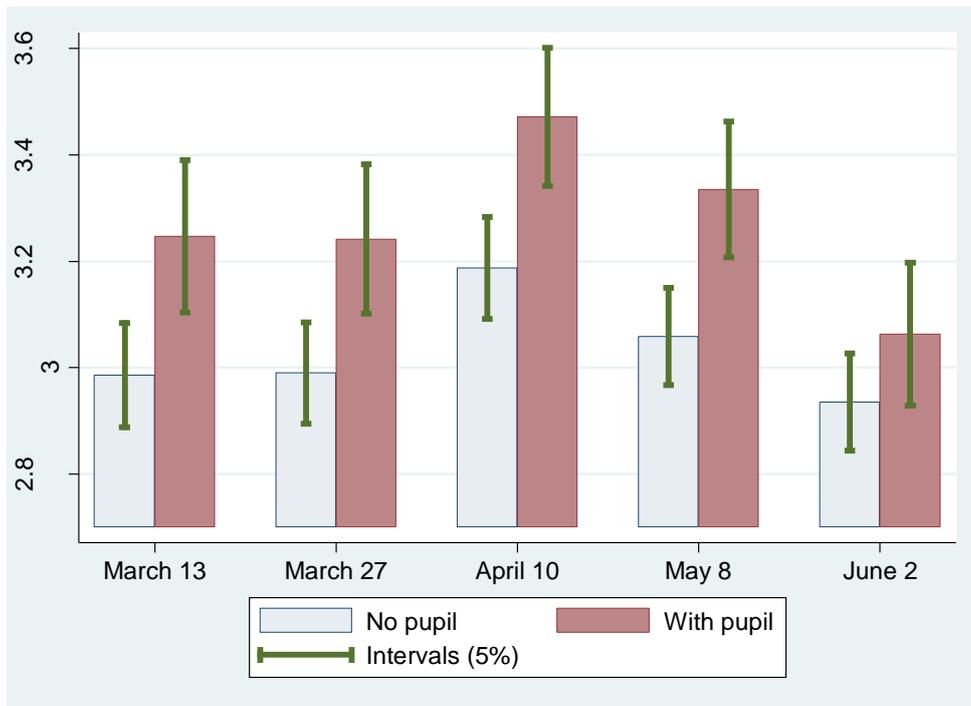

Anger

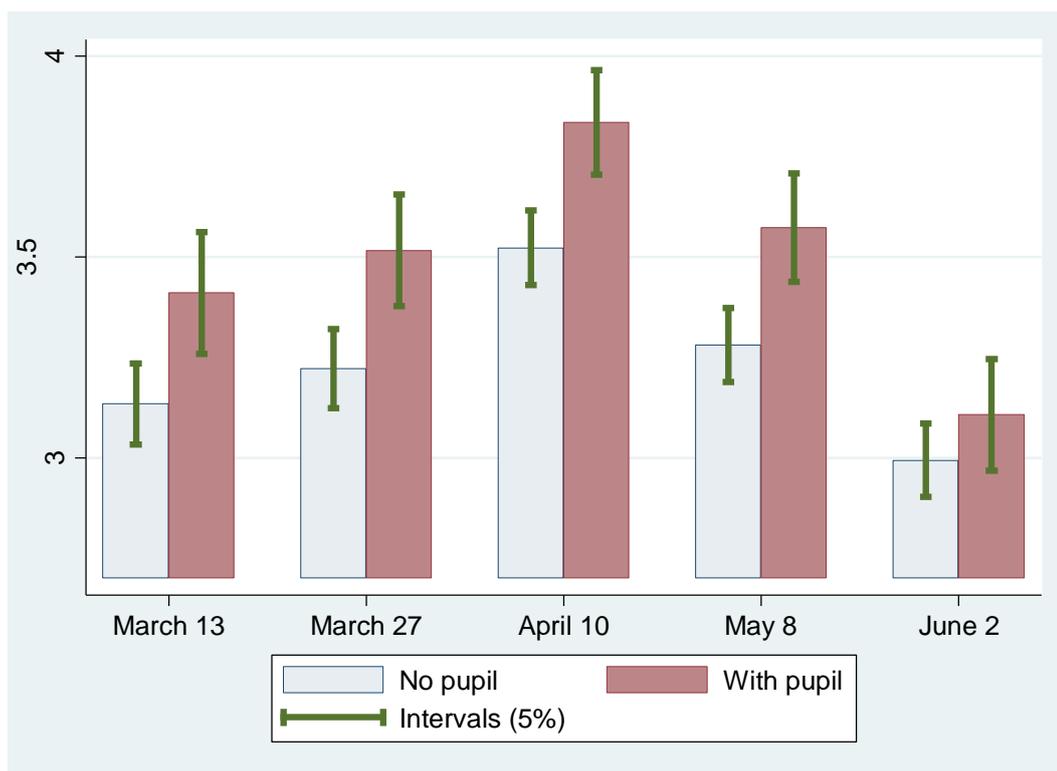

Fear

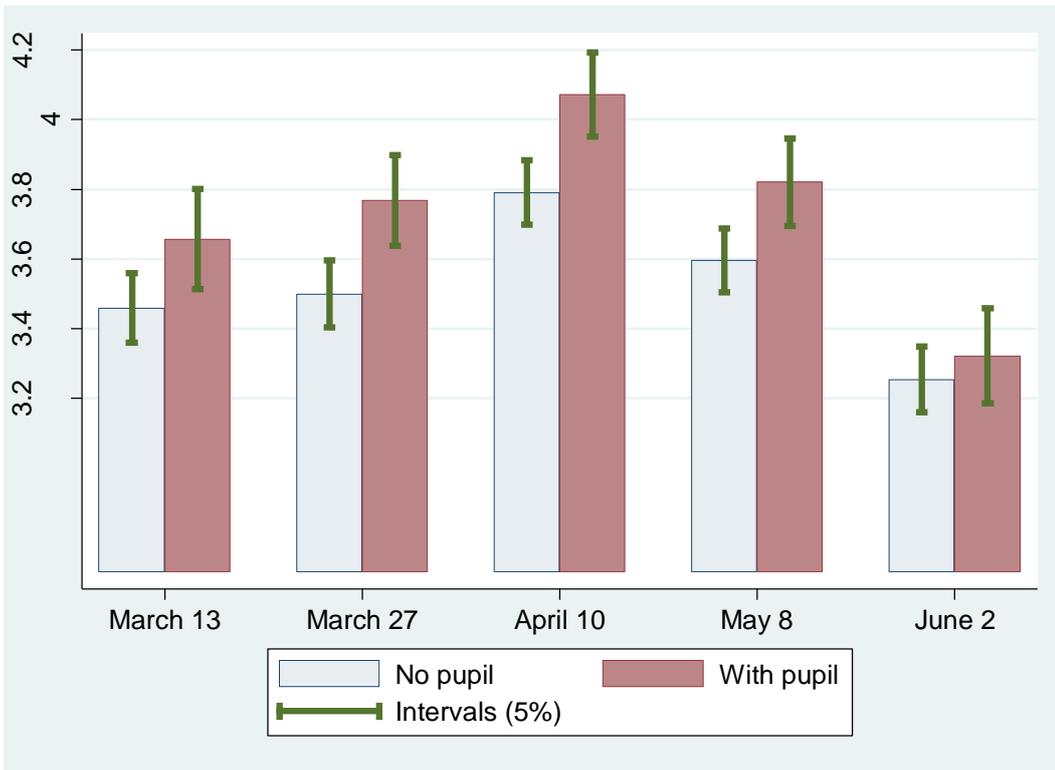

Anxiety

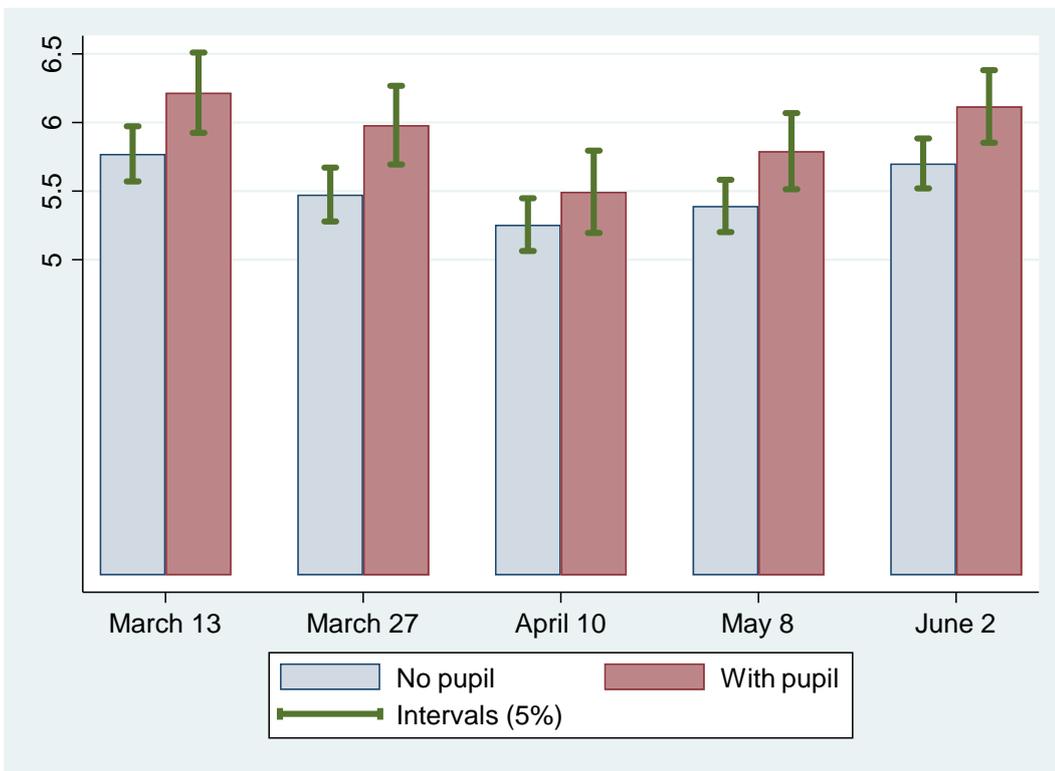

Happiness

Figure 4. Female

Table 1. Definitions of key variables and their basic statistics

|  | Definition | (1) Male | (2) Female |
|---|---|---|---|
| *Anger* | 'Within two weeks, how much have you felt the emotions of anger?' Please answer in a scale from 1 (I have not felt this emotion at all) to 5 (I have felt this emotion strongly). | 2.97 | 3.10 |
| *Fear* | 'Within two weeks, how much have you felt the emotions of fear?' Please answer in a scale from 1 (I have not felt this emotion at all) to 5 (I have felt this emotion strongly). | 3.00 | 3.30 |
| *Anxiety* | 'Within two weeks, how much have you felt the emotions of anxiety?' Please answer in a scale from 1 (I have not felt this emotion at all) to 5 (I have felt this emotion strongly). | 3.26 | 3.58 |
| *Happiness* | *'Now, to what degree are you currently feeling happiness? On a scale of 0 to 10, where 10 is 'very happy' and 0 is 'very unhappy,' How do you rate your current level?'.* | 5.04 | 5.62 |
| *Primary* | Equals 1 if respondent's child is primary school pupil, 0 otherwise | 0.21 | 0.29 |
| *Junior High* | Equals 1 if respondent's child is junior high school student, 0 otherwise | 0.11 | 0.14 |
| *School Closure A* | School closure rate in each prefecture. Its rates in the fifth wave are 0 in all prefectures (%) | 77.8 | 77.5 |
| *School Closure B* | School closure rate in each prefecture. Partial reopening in the fifth wave is considered as closed (%) | 85.4 | 85.1 |
| *Schooling Years* | Respondent's schooling years | 14.4 | 14.1 |
| *Income* | Respondent's annual household income. (Million yen) | 5.70 | 5.53 |
| *Ages* | Respondent's ages | 36.2 | 36.3 |
| *Wave 2* | Equals 1 if survey is second wave, 0 otherwise | 0.20 | 0.20 |
| *Wave 3* | Equals 1 if survey is third wave, 0 otherwise | 0.20 | 0.20 |
| *Wave 4* | Equals 1 if survey is fourth wave, 0 otherwise | 0.20 | 0.20 |
| *Wave 5* | Equals 1 if survey is third wave, 0 otherwise | 0.20 | 0.20 |

Note: Sample is limited to workers under 50 years old.

Table 2. Results of cross terms. Dependent variables: Mental conditions (Fixed effects model)

| | Male | | | | Female | | | |
|---|---|---|---|---|---|---|---|---|
| | (1) | (2) | (3) | (4) | (5) | (6) | (7) | (8) |
| | Anger | Fear | Anxiety | Happiness | Anger | Fear | Anxiety | Happiness |
| Wave 1 ×Primary | 0.11 (1.03) | 0.01 (0.06) | 0.03 (0.34) | 0.01 (0.07) | −0.24 (−0.26) | −0.02 (−0.25) | −0.07 (−0.77) | 0.22 (1.49) |
| Wave 2 ×Primary | −0.01 (−0.09) | −0.03 (−0.28) | −0.05 (−0.57) | −0.01 (−0.06) | −0.04 (−0.41) | −0.02 (−0.26) | −0.01 (−0.12) | 0.33** (2.43) |
| Wave 4 ×Primary | 0.01 (0.09) | 0.02 (0.25) | 0.03 (0.30) | −0.34** (−2.21) | −0.03 (−0.35) | −0.02 (−0.48) | −0.07 (−0.92) | 0.14 (1.01) |
| Wave 5 ×Primary | −0.05 (−0.49) | −0.04 (−0.48) | −0.12 (−1.34) | 0.04 (0.26) | −0.18** (−2.30) | −0.20** (−2.42) | −0.23** (−2.63) | 0.21 (1.41) |
| Wave 1 ×Junior High | 0.07 (0.51) | −0.02 (−0.16) | −0.03 (−0.18) | 0.24 (1.09) | 0.24* (1.66) | 0.40*** (3.10) | 0.12 (0.89) | −0.05 (−0.30) |
| Wave 2 ×Junior High | −0.07 (−0.48) | −0.13 (−0.92) | −0.03 (−0.24) | −0.01 (−0.06) | 0.18 (1.46) | 0.32*** (2.65) | 0.11 (0.85) | −0.03 (−0.22) |
| Wave 4 ×Junior High | −0.03 (−0.27) | −0.09 (−0.90) | −0.02 (−0.23) | 0.01 (0.06) | 0.07 (0.55) | 0.14 (1.30) | 0.12 (0.96) | 0.16 (0.76) |
| Wave 5 ×Junior High | −0.01 (−0.05) | −0.01 (−0.07) | 0.10 (0.75) | 0.26 (1.21) | 0.01 (0.07) | 0.25* (1.94) | 0.02 (0.15) | −0.04 (−0.19) |
| Wave 1 | −0.25*** (−4.87) | −0.46*** (−9.49) | −0.40*** (−7.92) | 0.33*** (4.01) | −0.21*** (−4.04) | −0.45*** (−8.42) | −0.36*** (−6.87) | 0.53*** (6.05) |
| Wave 2 | −0.21*** (−4.81) | −0.32*** (−6.96) | −0.28*** (−6.48) | 0.15** (2.32) | −0.21*** (−4.37) | −0.35*** (−7.29) | −0.32*** (−6.69) | 0.20** (2.59) |
| Wave 3 | <default> | | | | <default> | | | |
| Wave 4 | −0.05 (−1.18) | −0.17*** (−3.77) | −0.16*** (−3.58) | 0.11 (1.41) | −0.11** (−2.33) | −0.25*** (−5.52) | −0.21*** (−4.59) | 0.13* (1.86) |
| Wave 5 | −0.16*** (−3.33) | −0.36*** (−7.57) | −0.34*** (−7.27) | 0.18** (2.25) | −0.24*** (−4.26) | −0.57*** (−10.9) | −0.54*** (−10.7) | 0.48*** (5.94) |
| Within R-Square | 0.02 | 0.05 | 0.04 | 0.02 | 0.02 | 0.09 | 0.09 | 0.04 |
| Groups | 734 | 734 | 734 | 734 | 677 | 677 | 677 | 677 |
| Obs. | 3,670 | 3,670 | 3,670 | 3,670 | 3,385 | 3,385 | 3,385 | 3,385 |

Note: Numbers within parentheses are robust standard errors clustered on individuals. *** and ** indicate the statistical significance at 1% and 5% levels, respectively.

Table 3. Results of cross terms. Dependent variables: Mental conditions (Fixed effects model). Sub-sample of gradation from University at least (schooling years >=16).

|  | Male | | | | Female | | | |
|---|---|---|---|---|---|---|---|---|
|  | (1) *Anger* | (2) *Fear* | (3) *Anxiety* | (4) *Happiness* | (5) *Anger* | (6) *Fear* | (7) *Anxiety* | (8) *Happiness* |
| Wave 1 ×Primary | −0.03 (−0.23) | −0.11 (−0.87) | −0.07 (−0.50) | −0.15 (−0.70) | −0.14 (−0.94) | −0.08 (−0.61) | −0.02 (−0.17) | −0.08 (−0.35) |
| Wave 2 ×Primary | −0.18 (−1.45) | −0.17 (−1.33) | −0.18 (−1.52) | −0.05 (−0.25) | 0.01 (0.08) | 0.03 (0.27) | −0.002 (−0.02) | 0.05 (0.27) |
| Wave 4 ×Primary | −1.24 (−0.96) | −0.12 (−1.05) | −0.04 (−0.37) | −0.37* (−1.86) | −0.13 (−1.04) | −0.03 (−0.30) | −0.12 (−1.01) | 0.10 (0.51) |
| Wave 5 ×Primary | −0.19 (−1.44) | −0.20* (−1.68) | −0.23* (−1.89) | 0.01 (0.04) | −0.13 (−0.98) | −0.03 (−0.24) | −0.12 (−0.88) | 0.05 (0.25) |
| Wave 1 ×Junior High | 0.08 (0.33) | −0.39** (−2.09) | −0.15 (−0.72) | 0.40 (1.06) | 0.53 (1.45) | 0.21 (0.87) | 0.18 (0.76) | −0.10 (−0.35) |
| Wave 2 ×Junior High | −0.02 (−0.08) | −0.13 (−0.68) | −0.07 (−0.37) | 0.14 (0.59) | 0.60** (2.34) | 0.55** (2.30) | 0.32 (1.21) | −0.12 (−0.52) |
| Wave 4 ×Junior High | −0.08 (−0.53) | −0.08 (−0.57) | 0.01 (0.08) | 0.17 (0.51) | 0.23 (0.84) | 0.11 (0.39) | −0.24 (−0.77) | 0.47 (0.86) |
| Wave 5 ×Junior High | −0.02 (−0.19) | −0.01 (−0.06) | 0.12 (0.76) | 0.49 (1.48) | 0.06 (0.22) | 0.23 (1.02) | 0.06 (0.21) | 0.11 (0.28) |
| Wave 1 | −0.25*** (−3.47) | −0.47*** (−7.15) | −0.41*** (−5.87) | 0.37*** (3.51) | −0.22*** (−2.70) | −0.42*** (−5.41) | −0.41*** (−4.94) | 0.54*** (4.08) |
| Wave 2 | −0.10 (−1.60) | −0.24*** (−3.77) | −0.21*** (−3.28) | 0.22** (2.42) | −0.19*** (−2.70) | −0.33*** (−4.97) | −0.32*** (−4.17) | 0.19 (1.61) |
| Wave 3 | | <default> | | | | <default> | | |
| Wave 4 | 0.03 (0.52) | −0.13** (−2.10) | −0.15** (−2.48) | 0.03 (0.33) | −0.10 (−1.40) | −0.23*** (−3.39) | −0.22*** (−2.93) | 0.18 (1.54) |
| Wave 5 | −0.09 (−1.42) | −0.35*** (−5.52) | −0.38*** (−5.76) | 0.09 (0.95) | −0.24*** (−3.01) | −0.63*** (−8.54) | −0.57*** (−6.97) | 0.51*** (3.87) |
| Within R-Square | 0.02 | 0.08 | 0.06 | 0.02 | 0.03 | 0.10 | 0.09 | 0.04 |
| Groups | 385 | 385 | 385 | 385 | 270 | 677 | 677 | 677 |
| Obs. | 1,925 | 1,925 | 1,925 | 1,925 | 1,350 | 3,385 | 3,385 | 3,385 |

Note: Numbers within parentheses are robust standard errors clustered on individuals. *** and ** indicate the statistical significance at 1% and 5% levels, respectively.

Table 4. Results of cross terms. Dependent variables: Mental conditions (Fixed effects model). Sub-sample of graduation from high school or college at most (schooling years <16).

|  | Male | | | | Female | | | |
|---|---|---|---|---|---|---|---|---|
|  | (1) Anger | (2) Fear | (3) Anxiety | (4) Happiness | (5) Anger | (6) Fear | (7) Anxiety | (8) Happiness |
| Wave 1 ×Primary | 0.28 (1.57) | 0.17 (1.01) | 0.18 (1.14) | 0.25 (0.95) | 0.03 (0.21) | 0.01 (0.07) | −0.11 (−0.89) | 0.45** (2.34) |
| Wave 2 ×Primary | 0.15 (1.07) | 0.13 (0.89) | 0.10 (0.83) | 0.03 (0.13) | −0.07 (−0.60) | −0.03 (−0.27) | 0.004 (0.04) | 0.55*** (2.92) |
| Wave 4 ×Primary | 0.10 (0.64) | 0.18 (1.28) | 0.09 (0.70) | −0.28 (−1.18) | 0.01 (0.11) | −0.04 (−0.42) | −0.06 (−0.55) | 0.22 (1.14) |
| Wave 5 ×Primary | 0.08 (0.50) | 0.15 (1.00) | 0.01 (0.08) | 0.10 (0.40) | −0.21* (−1.76) | −0.30** (−2.54) | −0.30** (−2.50) | 0.31 (1.55) |
| Wave 1 ×Junior High | 0.08 (0.41) | 0.20 (0.93) | 0.02 (0.09) | 0.14 (0.54) | 0.10 (0.63) | 0.42*** (2.70) | 0.07 (0.46) | −0.02 (−0.10) |
| Wave 2 ×Junior High | −0.06 (−0.31) | −0.14 (−0.67) | 0.01 (0.04) | −0.08 (−0.34) | 0.05 (0.38) | 0.26* (1.77) | 0.05 (0.36) | 0.01 (0.06) |
| Wave 4 ×Junior High | 0.03 (0.15) | −0.14 (−0.90) | −0.08 (−0.52) | −0.15 (−0.51) | −0.02 (−0.15) | 0.16 (1.36) | 0.23* (1.82) | 0.10 (0.43) |
| Wave 5 ×Junior High | 0.04 (0.24) | −0.05 (−0.22) | 0.04 (0.22) | 0.06 (0.20) | −0.04 (−0.25) | 0.21 (1.38) | 0.001 (0.01) | −0.15 (−0.54) |
| Wave 1 | −0.28*** (−3.77) | −0.43*** (−5.95) | −0.37*** (−5.01) | 0.22* (1.75) | −0.18** (−2.56) | −0.45*** (−5.99) | −0.31*** (−4.45) | 0.52*** (4.29) |
| Wave 2 | −0.38*** (−5.38) | −0.38*** (−5.99) | −0.37*** (−5.94) | 0.05 (0.56) | −0.23*** (−3.29) | −0.37*** (−5.28) | −0.33*** (−5.31) | 0.20* (1.90) |
| Wave 3 | <default> | | | | <default> | | | |
| Wave 4 | −0.13* (−1.94) | −0.19*** (−2.70) | −0.14** (−2.20) | 0.19* (1.68) | −0.12* (−1.67) | −0.27*** (−4.38) | −0.19*** (−3.40) | 0.10 (1.09) |
| Wave 5 | −0.26*** (−3.44) | −0.36*** (−5.11) | −0.30*** (−4.42) | 0.25** (2.07) | −0.22*** (−2.85) | −0.54*** (−7.31) | −0.53*** (−7.57) | 0.47*** (4.54) |
| Within R-Square | 0.03 | 0.05 | 0.04 | 0.02 | 0.03 | 0.09 | 0.10 | 0.05 |
| Groups | 328 | 328 | 328 | 328 | 391 | 391 | 391 | 391 |
| Obs. | 1,640 | 1,640 | 1,640 | 1,640 | 1,955 | 1,955 | 1,955 | 1,955 |

Note: Numbers within parentheses are robust standard errors clustered on individuals. *** and ** indicate the statistical significance at 1% and 5% levels, respectively.

Table 5. Results of school closure rate. Dependent variables: Mental conditions (Fixed effects model): Male sample

|  | Specification A | | | | Specification B | | | |
| --- | --- | --- | --- | --- | --- | --- | --- | --- |
|  | (1) *Anger* | (2) *Fear* | (3) *Anxiety* | (4) *Happiness* | (5) *Anger* | (6) *Fear* | (7) *Anxiety* | (8) *Happiness* |
| School Closure A ×Primary | 1.09 (1.40) | 0.56 (0.80) | 1.27 (1.64) | −0.99 (−0.83) | | | | |
| School Closure A ×Junior High | −0.17 (−0.17) | −0.76 (−0.71) | −1.45 (−1.28) | −1.53 (−1.09) | | | | |
| School Closure A | 0.31 (0.27) | 1.97* (1.66) | 2.26* (1.79) | −0.33 (−0.17) | | | | |
| School Closure B ×Primary | | | | | 1.61 (1.53) | 0.83 (0.77) | 1.67 (1.53) | −1.46 (−0.91) |
| School Closure B ×Junior High | | | | | −0.28 (−0.21) | −0.60 (−0.41) | −0.72 (−0.46) | −2.55 (−1.37) |
| School Closure B | | | | | 0.35 (0.33) | 0.99 (0.88) | 0.47 (0.41) | 1.59 (0.94) |
| Within R-Square | 0.02 | 0.05 | 0.04 | 0.01 | 0.02 | 0.05 | 0.04 | 0.01 |
| Groups | 734 | 734 | 734 | 734 | 677 | 677 | 677 | 677 |
| Obs. | 3,670 | 3,670 | 3,670 | 3,670 | 3,385 | 3,385 | 3,385 | 3,385 |

Note: In all columns, wave dummies are included, but the results are not reported. The numbers within parentheses are robust standard errors clustered on individuals.

For convenience of interpretation, coefficients are multiplied by 1000. *** and ** indicate the statistical significance at 1% and 5% levels, respectively.

Table 6. Results of school closure rate. Dependent variables: Mental conditions (Fixed effects model): Female sample

|  | Specification A | | | | Specification B | | | |
|---|---|---|---|---|---|---|---|---|
|  | (1) Anger | (2) Fear | (3) Anxiety | (4) Happiness | (5) Anger | (6) Fear | (7) Anxiety | (8) Happiness |
| School Closure A ×Primary | 1.48** (2.06) | 1.73** (2.52) | 1.85*** (2.64) | −0.68 (−0.58) | | | | |
| School Closure A ×Junior High | 1.08 (1.13) | −0.38 (−0.40) | 0.68 (0.65) | 0.18 (0.10) | | | | |
| School Closure A | 0.31 (0.25) | 0.69 (0.59) | −0.34 (−0.28) | −0.20 (−0.11) | | | | |
| School Closure B ×Primary | | | | | 2.22** (2.20) | 2.89** (2.17) | 1.11*** (2.62) | 0.35 (0.20) |
| School Closure B ×Junior High | | | | | 0.95 (0.69) | −0.73 (−0.51) | 0.10 (0.07) | 0.37 (0.13) |
| School Closure B | | | | | −0.72 (−0.74) | −1.10 (−1.06) | −1.10 (−1.10) | 2.83* (−1.80) |
| Within R-Square | 0.02 | 0.08 | 0.09 | 0.04 | 0.01 | 0.08 | 0.09 | 0.04 |
| Groups | 677 | 677 | 677 | 677 | 677 | 677 | 677 | 677 |
| Obs. | 3,385 | 3,385 | 3,385 | 3,385 | 3,385 | 3,385 | 3,385 | 3,385 |

Note: In all columns, wave dummies are included, but the results are not reported. The numbers within parentheses are robust standard errors clustered on individuals.

For convenience of interpretation, coefficients are multiplied by 1000. *** and ** indicate the statistical significance at 1% and 5% levels, respectively.

Table 7. Results of school closure rate. Dependent variables: Mental conditions (Fixed effects model): Sub-sample of graduation from university at least (schooling years >=16).

Male sample.

|  | Specification A | | | | Specification B | | | |
| --- | --- | --- | --- | --- | --- | --- | --- | --- |
|  | (1) Anger | (2) Fear | (3) Anxiety | (4) Happiness | (5) Anger | (6) Fear | (7) Anxiety | (8) Happiness |
| School Closure A ×Primary | 1.33 (1.33) | 1.29 (1.56) | 1.69* (1.79) | −1.14 (−0.81) | | | | |
| School Closure A ×Junior High | 0.18 (0.13) | −1.47 (−1.10) | −1.91 (−1.24) | −2.61 (−1.25) | | | | |
| School Closure A | 2.17 (1.34) | 2.15 (1.34) | −0.18 (−0.10) | −1.78 (−1.25) | | | | |
| School Closure B ×Primary | | | | | 1.76 (1.27) | 2.00 (1.50) | 2.59* (1.84) | −1.24 (−0.56) |
| School Closure B ×Junior High | | | | | 0.15 (0.09) | −1.68 (−1.04) | −1.26 (−0.64) | −3.63 (−1.21) |
| School Closure B | | | | | 1.47 (1.02) | 0.74 (0.58) | −1.58 (−0.64) | −1.09 (−0.50) |
| Within R-Square | 0.02 | 0.07 | 0.05 | 0.02 | 0.02 | 0.07 | 0.05 | 0.02 |
| Groups | 406 | 406 | 406 | 406 | 406 | 406 | 406 | 406 |
| Obs. | 2,030 | 2,030 | 2,030 | 2,030 | 2,030 | 2,030 | 2,030 | 2,030 |

Note: In all columns, wave dummies are included, but the results are not reported. The numbers within parentheses are robust standard errors clustered on individuals.

For convenience of interpretation, coefficients are multiplied by 1000. *** and ** indicate the statistical significance at 1% and 5% levels, respectively.

Table 8. Results of school closure rate. Dependent variables: Mental conditions (Fixed effects model): Sub-sample of graduation from University at least (schooling years >=16).

Female sample

|  | Specification A | | | | Specification B | | | |
| --- | --- | --- | --- | --- | --- | --- | --- | --- |
|  | (1) *Anger* | (2) *Fear* | (3) *Anxiety* | (4) *Happiness* | (5) *Anger* | (6) *Fear* | (7) *Anxiety* | (8) *Happiness* |
| *School Closure A ×Primary* | 0.82 (0.43) | 0.36 (0.34) | 0.94 (0.87) | −0.78 (−0.47) | | | | |
| *School Closure A ×Junior High* | 2.71 (1.54) | −0.29 (−0.15) | 0.41 (0.21) | −3.25 (−1.25) | | | | |
| *School Closure A* | 1.59 (0.64) | 1.00 (0.43) | 1.44 (0.54) | −1.89 (−0.54) | | | | |
| *School Closure B ×Primary* | | | | | 1.33 (0.78) | 0.14 (0.08) | 1.21 (0.72) | 0.62 (0.26) |
| *School Closure B ×Junior High* | | | | | 5.22* (1.84) | 1.31 (0.38) | 1.89 (0.56) | −5.79 (−1.47) |
| *School Closure B* | | | | | −0.25 (−0.12) | −0.02 (−0.01) | 1.72 (0.94) | 1.33 (0.48) |
| Within R-Square | 0.02 | 0.09 | 0.08 | 0.04 | 0.02 | 0.09 | 0.08 | 0.04 |
| Groups | 286 | 286 | 286 | 286 | 286 | 286 | 286 | 286 |
| Obs. | 1,430 | 1,430 | 1,430 | 1,430 | 1,430 | 1,430 | 1,430 | 1,430 |

Note: In all columns, wave dummies are included, but the results are not reported. The numbers within parentheses are robust standard errors clustered on individuals.

For convenience of interpretation, coefficients are multiplied by 1000. *** and ** indicate the statistical significance at 1% and 5% levels, respectively.

Table 9. Results of school closure rate. Dependent variables: Mental conditions (Fixed effects model): Sub-sample of graduation from high school or college at most (schooling years <16).

Male sample.

|  | Specification A | | | | Specification B | | | |
|---|---|---|---|---|---|---|---|---|
|  | (1) Anger | (2) Fear | (3) Anxiety | (4) Happiness | (5) Anger | (6) Fear | (7) Anxiety | (8) Happiness |
| School Closure A ×Primary | 0.80 (0.97) | −0.38 (−0.32) | 0.72 (0.56) | −0.86 (−0.43) | | | | |
| School Closure A ×Junior High | −0.40 (−0.28) | −0.02 (−0.01) | −0.89 (−0.54) | −0.33 (−0.17) | | | | |
| School Closure A | −1.10 (−0.68) | 1.69 (0.97) | 4.18** (2.11) | 1.21 (0.41) | | | | |
| School Closure B ×Primary | | | | | 1.49 (0.91) | −0.50 (−0.29) | 0.66 (0.39) | −1.82 (−0.79) |
| School Closure B ×Junior High | | | | | −0.58 (−0.29) | 0.53 (0.23) | 0.03 (0.01) | −1.27 (−0.55) |
| School Closure B | | | | | −0.81 (−0.50) | 1.27 (0.70) | 2.73 (1.60) | 4.94* (1.89) |
| Within R-Square | 0.03 | 0.04 | 0.04 | 0.01 | 0.03 | 0.04 | 0.04 | 0.01 |
| Groups | 328 | 328 | 328 | 328 | 328 | 328 | 328 | 328 |
| Obs. | 1,640 | 1,640 | 1,640 | 1,640 | 1,640 | 1,640 | 1,640 | 1,640 |

Note: In all columns, wave dummies are included, but the results are not reported. The numbers within parentheses are robust standard errors clustered on individuals.

For convenience of interpretation, coefficients are multiplied by 1000. *** and ** indicate the statistical significance at 1% and 5% levels, respectively.

Table 10. Results of school closure rate. Dependent variables: Mental conditions (Fixed effects model): Sub-sample of graduation from high school or college at most (schooling years <16).

Female sample

|  | Specification A | | | | Specification B | | | |
| --- | --- | --- | --- | --- | --- | --- | --- | --- |
|  | (1) Anger | (2) Fear | (3) Anxiety | (4) Happiness | (5) Anger | (6) Fear | (7) Anxiety | (8) Happiness |
| School Closure A ×Primary | 1.92** (2.05) | 2.75*** (2.94) | 2.41*** (2.62) | −0.57 (−0.35) | | | | |
| School Closure A ×Junior High | 0.46 (0.40) | −0.39 (−0.35) | 0.61 (0.50) | 1.33 (0.57) | | | | |
| School Closure A | −0.001 (−0.07) | 0.47 (0.34) | −0.70 (−0.54) | −0.01 (−0.03) | | | | |
| School Closure B ×Primary | | | | | 2.68** (2.14) | 3.49*** (2.62) | 3.31** (2.58) | 0.34 (0.14) |
| School Closure B ×Junior High | | | | | −0.11 (−0.07) | −1.21 (−0.78) | −0.37 (−0.27) | 1.59 (0.45) |
| School Closure B | | | | | −0.94 (−0.79) | −1.59 (−1.29) | −2.44** (−2.06) | 3.09 (1.62) |
| Within R-Square | 0.02 | 0.08 | 0.10 | 0.05 | 0.02 | 0.08 | 0.10 | 0.05 |
| Groups | 391 | 391 | 391 | 391 | 391 | 391 | 391 | 391 |
| Obs. | 1,955 | 1,955 | 1,955 | 1,955 | 1,955 | 1,955 | 1,955 | 1,955 |

Note: In all columns, wave dummies are included, but the results are not reported. The numbers within parentheses are robust standard errors clustered on individuals.

For convenience of interpretation, coefficients are multiplied by 1000. *** and ** indicate the statistical significance at 1% and 5% levels, respectively.

# Appendix

Table 11. Results of cross terms. Dependent variables: Mental conditions (Random effects Ordered-Probit model)

|  | Male | | | | Female | | | |
|---|---|---|---|---|---|---|---|---|
|  | (1) *Anger* | (2) *Fear* | (3) *Anxiety* | (4) *Happiness* | (5) *Anger* | (6) *Fear* | (7) *Anxiety* | (8) *Happiness* |
| Wave 1 ×Primary | 0.16 (1.03) | −0.01 (−0.05) | 0.02 (0.11) | −0.003 (−0.02) | −0.04 (−0.29) | −0.07 (−0.51) | −0.15 (−1.04) | 0.18 (1.42) |
| Wave 2 ×Primary | −0.01 (−0.11) | −0.05 (−0.35) | −0.09 (−0.72) | −0.02 (−0.13) | −0.05 (−0.40) | −0.07 (−0.53) | −0.05 (−0.41) | 0.29** (2.41) |
| Wave 4 ×Primary | 0.01 (0.06) | 0.02 (0.13) | 0.02 (0.15) | −0.15 (−1.20) | −0.04 (−0.34) | −0.07 (−0.62) | −0.14 (−1.09) | 0.11 (0.92) |
| Wave 5 ×Primary | −0.07 (−0.48) | −0.07 (−0.54) | −0.19 (−1.41) | 0.01 (0.08) | −0.26** (−2.03) | −0.34*** (−2.66) | −0.40*** (−2.84) | 0.18 (1.39) |
| Wave 1 ×Junior High | 0.10 (0.48) | −0.03 (−0.12) | −0.03 (−0.12) | 0.09 (0.53) | 0.34* (1.69) | 0.60*** (3.02) | 0.19 (0.93) | −0.04 (−0.27) |
| Wave 2 ×Junior High | −0.09 (−0.49) | −0.18 (−0.89) | −0.04 (−0.20) | −0.02 (−0.09) | 0.25 (1.45) | 0.48** (2.55) | 0.18 (0.94) | −0.04 (−0.33) |
| Wave 4 ×Junior High | −0.04 (−0.26) | −0.14 (−0.88) | −0.03 (−0.19) | −0.01 (−0.04) | 0.09 (0.53) | 0.21 (1.29) | 0.19 (1.01) | 0.14 (0.74) |
| Wave 5 ×Junior High | −0.02 (−0.11) | 0.01 (0.05) | 0.15 (0.80) | 0.10 (0.55) | 0.01 (0.08) | 0.38* (1.96) | 0.06 (0.30) | −0.02 (−0.13) |
| Wave 1 | −0.34*** (−4.80) | −0.66*** (−9.32) | −0.57*** (−7.69) | 0.15** (2.44) | −0.30*** (−4.04) | −0.69*** (−8.22) | −0.56*** (−6.79) | 0.48*** (6.29) |
| Wave 2 | −0.30*** (−4.78) | −0.45*** (−6.93) | −0.41*** (−6.33) | 0.07 (1.15) | −0.31*** (−4.37) | −0.53*** (−7.11) | −0.52*** (−6.77) | 0.18*** (2.63) |
| Wave 3 | <default> | | | | <default> | | | |
| Wave 4 | −0.07 (−1.21) | −0.24*** (−3.77) | −0.23*** (−3.53) | 0.05 (0.78) | −0.16** (−2.39) | −0.39*** (−5.46) | −0.34*** (−4.65) | 0.12* (1.89) |
| Wave 5 | −0.22*** (−3.32) | −0.53*** (−7.50) | −0.51*** (−7.19) | 0.07 (1.19) | −0.34*** (−4.30) | −0.86*** (−10.5) | −0.85*** (−10.2) | 0.42*** (5.99) |
| Wald Chi-Square | 53 | 160 | 140 | 123 | 84 | 248 | 248 | 128 |
| Groups | 734 | 734 | 734 | 734 | 734 | 734 | 734 | 734 |
| Obs. | 3,670 | 3,670 | 3,670 | 3,670 | 3,670 | 3,670 | 3,670 | 3,670 |

Note: Numbers within parentheses are robust standard errors clustered on individuals. *** and ** indicate the statistical significance at 1% and 5% levels, respectively.

Table 12. Results of school closure rate. Dependent variables: Mental conditions (Random effects Ordered-Probit model): Male sample

|  | Specification A | | | | Specification B | | | |
| --- | --- | --- | --- | --- | --- | --- | --- | --- |
|  | (1) Anger | (2) Fear | (3) Anxiety | (4) Happiness | (5) Anger | (6) Fear | (7) Anxiety | (8) Happiness |
| School Closure A ×Primary | 1.42 (1.32) | 0.74 (0.73) | 1.68 (1.55) | −0.23 (−0.25) | | | | |
| School Closure A ×Junior High | −0.38 (−0.28) | −1.45 (−0.92) | −2.33 (−1.42) | 0.21 (0.17) | | | | |
| School Closure A | 1.14 (0.72) | 3.38* (1.95) | 4.01** (2.32) | −3.62** (−2.47) | | | | |
| School Closure B ×Primary | | | | | 2.13 (1.47) | 0.99 (0.63) | 2.11 (1.39) | −0.21 (−0.16) |
| School Closure B ×Junior High | | | | | −1.01 (−0.55) | −1.66 (−0.78) | −1.61 (−0.71) | 0.28 (0.23) |
| School Closure B | | | | | 1.14 (0.81) | 2.03 (1.31) | 1.51 (0.98) | −1.83 (−1.29) |
| Wald Chi-Square | 50 | 159 | 135 | 136 | 50 | 159 | 135 | 136 |
| Groups | 734 | 734 | 734 | 734 | 734 | 734 | 734 | 734 |
| Obs. | 3,670 | 3,670 | 3,670 | 3,670 | 3,670 | 3,670 | 3,670 | 3,670 |

Note: In all columns, wave dummies are included, but the results are not reported. The numbers within parentheses are robust standard errors clustered on individuals.

For convenience of interpretation, coefficients are multiplied by 1000. *** and ** indicate the statistical significance at 1% and 5% levels, respectively.

Table 13. Results of school closure rate. Dependent variables: Mental conditions (Random effects Ordered-Probit model): Female sample

|  | Specification A | | | | Specification B | | | |
| --- | --- | --- | --- | --- | --- | --- | --- | --- |
|  | (1) | (2) | (3) | (4) | (5) | (6) | (7) | (8) |
|  | *Anger* | *Fear* | *Anxiety* | *Happiness* | *Anger* | *Fear* | *Anxiety* | *Happiness* |
| School Closure A ×Primary | 2.26** | 2.94*** | 3.00*** | −0.68 | | | | |
| | (2.23) | (2.52) | (2.93) | (−0.65) | | | | |
| School Closure A ×Junior High | 1.52 | −0.74 | 0.71 | 0.01 | | | | |
| | (1.14) | (−0.52) | (0.47) | (0.05) | | | | |
| School Closure A | 0.11 | 0.04 | −0.59 | −0.24 | | | | |
| | (0.06) | (0.03) | (−0.34) | (−0.14) | | | | |
| School Closure B ×Primary | | | | | 3.48** | 4.07*** | 4.41*** | 0.21 |
| | | | | | (2.48) | (2.65) | (2.96) | (0.13) |
| School Closure B ×Junior High | | | | | 1.25 | −1.45 | −0.35 | 0.41 |
| | | | | | (0.68) | (−0.69) | (−0.63) | (0.17) |
| School Closure B | | | | | −1.27 | −2.23 | −1.71 | 2.16 |
| | | | | | (−0.91) | (−1.45) | (−1.21) | (1.58) |
| Wald Chi-Square | 79 | 228 | 240 | 119 | 79 | 228 | 240 | 119 |
| Groups | 677 | 677 | 677 | 677 | 677 | 677 | 677 | 677 |
| Obs. | 3,385 | 3,385 | 3,385 | 3,385 | 3,385 | 3,385 | 3,385 | 3,385 |

Note: In all columns, wave dummies are included, but the results are not reported. The numbers within parentheses are robust standard errors clustered on individuals.

For convenience of interpretation, coefficients are multiplied by 1000. *** and ** indicate the statistical significance at 1% and 5% levels, respectively.